\documentclass[review]{elsarticle}

\usepackage{lineno,hyperref}
\modulolinenumbers[5]

\journal{Journal of Computational Physics}

\usepackage{graphicx}
\usepackage{amsmath,amssymb}
\usepackage{amsthm}
\usepackage{xcolor}
\usepackage{mathrsfs}
\usepackage{paralist}
\usepackage{multirow}
\usepackage[T1]{fontenc}
\usepackage[utf8]{inputenc}
\usepackage[english]{babel}
\usepackage{subfigure}
\DeclareMathAlphabet\mathbfcal{OMS}{cmsy}{b}{n}
\usepackage{caption}
\usepackage{lineno}
\usepackage{bbm}
\usepackage[normalem]{ulem}
\usepackage[percent]{overpic}
\usepackage{cleveref}
\usepackage{color}
\usepackage{tikz}
\usepackage{yhmath}

\newcommand{\ud}{\mathrm{d}}

\newcommand{\ie}{\emph{i.e. }} 

\newcommand{\pd}[2]{\frac{\partial{#1}}{\partial{#2}}} 
\newcommand{\der}[2]{\frac{\ud{#1}}{\ud{#2}}} 

\newcommand{\del}[1]{}

\newcommand{\greenline}{\raisebox{2pt}{\tikz{\draw[-,black!40!green,solid,line width = 1.5pt](0,0) -- (5mm,0);}}}
\newcommand{\redline}{\raisebox{2pt}{\tikz{\draw[-,black!1!red,solid,line width = 1.5pt](0,0) -- (5mm,0);}}}

\usepackage[a4paper,bindingoffset=0.0in,%
            left=1.0in,right=1.0in,top=1.in,bottom=1.in,%
            footskip=.4in]{geometry}

\begin{document}

\begin{frontmatter}

\title{A high-order diffused-interface approach for two-phase compressible flow simulations using a Discontinuous Galerkin framework}

\author[mymainaddress]{Niccol{\`o} Tonicello\corref{mycorrespondingauthor}}
\cortext[mycorrespondingauthor]{Corresponding author}
\ead{tonicnic@stanford.edu}
\author[mymainaddress]{Matthias Ihme}

\address[mymainaddress]{Department of Mechanical Engineering, Stanford University, Stanford, CA 94305, United States}

\begin{abstract}
A diffused-interface approach based on the Allen-Cahn phase field equation is developed within a high-order Discontinuous Galerkin framework. The interface capturing technique is based on the balance between explicit diffusion and sharpening terms in the phase field equation, where the former term involves the computation of the local interface normal vectors. Due to the well-known Gibbs phenomenon encountered in high-order discretisations of steep profiles such as shocks and/or interfaces, the accurate evaluation of the normal vector requires special consideration. To this end, a non-linear preconditioning strategy is proposed in this work where an additional smooth level-set function advected by the velocity field is used for the evaluation of the normal vectors. It is shown that for appropriate choices of numerical fluxes and parameters of the model, the phase field remains bounded without any need for explicit regularisation. The proposed diffused-interface technique is implemented within a five equation model for fully compressible two-phase flows. In order to preserve isolated interfaces, a quasi-conservative discretisation of the five equation model is employed. A series of numerical experiments of increasing complexity are performed in order to assess the accuracy and robustness of the developed methodology, including two-phase flows involving viscous effects, gravitational forces, and surface tension.
\end{abstract}

\begin{keyword}
High-order methods, Discontinuous Galerkin, phase field, two-phase flows
\end{keyword}

\end{frontmatter}

\section{Introduction}
The constant increase in computational power has made the use of Computational Fluid Dynamics (CFD) an increasingly valuable tool in the design process of complex industrial applications \cite{slotnick2014cfd}. Commonly available commercial CFD softwares are mostly based on low-order numerical schemes such as finite volumes or finite differences methods. While these numerical methods are particularly robust and reliable, they usually lack in accuracy for complex applications. Along these lines, the development of innovative high-order schemes, such as Discontinuous Galerkin \cite{hesthaven:book,cockburn:98,cockburn:98b,CANTWELL2015205}, Flux Reconstruction \cite{huynh2007flux,vincent2011new,witherden2014pyfr} and Spectral Difference \cite{kopriva1996conservative,karniadakis2013spectral} methods, has experienced a significant growth in the CFD community, representing a promising alternative for the next generation of CFD commercial codes.

Whereas the use of high-order spectral element methods is becoming more common in the simulation of compressible aerodynamics problems \cite{bosnyakov2014high,lv2014discontinuous,mengaldo2021industry,tonicello2022analysis,tonicello2022turbulence,ching2019shock,moxey2020nektar++,lv2021discontinuous,dzanic2022positivity,ferrer2023high}, multiphase applications still largely rely on low-order discretisations, in particular for fully compressible flows. 

High-order numerical schemes, due to the low numerical dissipation and dispersion errors \cite{lele1992compact,bogey2004family,van2008stability,vincent2011insights,moura2015linear,vanharen2017revisiting,mengaldo2018spatial,tonicello2021comparative} can be beneficial in the resolution of small-scale structures which are often encounter in the simulation of multiphase flows. In contrast, because of the delicate nature of such schemes, careful attention is needed in order to retain stability properties.

One common approach in the simulation of two-phase flows relies on the concept of volume fraction and its most popular formulation is commonly known as Volume of Fluid (VOF) method \cite{debar1974fundamentals,nichols1975methods}. The volume fraction is an auxiliary function that is advected by the velocity field which varies between zero and one and represents the ratio of primary to secondary fluid at a given computational grid point. The interface between two immiscible fluids is then represented by a sharp variation of the volume fraction. 

The phase field approach \cite{olsson2005conservative,sun2007sharp,chiu2011conservative} follows a similar concept with respect to VOF methodologies where a function bounded between zero and one is used to distinguish the two phases. However, instead of relying on the explicit reconstruction of the interface as it is custom in VOF methods, a balance between a diffusion and an anti-diffusion term in the transport equation of the phase field is used in order to maintain a sharp, but at the same time sufficiently smooth, diffused profile of the interface. 

Once a clear separation between the two phases is identified, it is possible to introduce an additional set of equations that can be used to model each phase, potentially, with considerably different transport properties and equations of state. At the same time, localised forces or fluxes can be easily imposed at the interface as it often happens in the modelling of surface tension effects \cite{brackbill1992continuum}.

The objective of the present work is to develop a simple but, at the same time, accurate and robust approach to deal with compressible two-phase flows within the framework of the Discontinuous Galerkin method. The proposed methodology is able to preserve important properties in the simulation of two-phase flows such as boundedness of the volume fraction, accurate evaluation of the interface normal vectors, low mass conservation errors and exact resolution of contact discontinuities. From an implementational point of view, the proposed approach does not need explicit limiting, interface reconstruction or mass redistribution, but it is simply based on appropriate definitions of numerical fluxes and additional equations. Finally, the high-order spatial discretisation provides significant advantages from many different points of view, from considerably reducing mass conservation errors, to avoiding spurious deformations of the interface for long time integration. Along the same lines, the generality of this method allows easy implementation of different physical phenomena such as viscous, gravitational and surface tension forces. In other words, the novelty of the present method resides in its generality and flexibility: with only minor, targeted modifications on the central kernels of the numerical scheme, it is possible to preserve a large amount of desirable features in two-phase flows simulations, adding, at the same time, the benefits given by the high-order spatial discretisation.

The paper is structured as follows. In section \ref{sec:5eq}, the five equation model for two immiscible compressible fluids is introduced, including all the additional terms and equations associated to the specific interface capturing approach employed in the present work. In section \ref{sec:DG}, the Discontinuous Galerkin method is first outlined in its general formulation for conservation laws. Its implementation for the specific case of the five equation model is then presented, including additional details on the interface capturing technique, the specific choices of numerical fluxes and their related properties. Subsequently, section \ref{sec:results} is dedicated to the numerical results in which both kinematics tests and fully coupled two-phase flows for increasing levels of complexity are considered in order to validate the present implementation. Within the benchmark cases herein considered, extensive studies on different types of elements, different orders of approximation and grid-convergence investigations are carried out in order to assess the robustness and accuracy of the proposed methodology for a wide range of different problems in two-phase flows. Finally, in section \ref{sec:conclusions} the key conclusions of this work are discussed.
\section{Five equation model} \label{sec:5eq}
The system of conservation laws considered in this work aims at modelling two-phase compressible flows including viscous effects, gravitational forces and surface tension. The five equation model \cite{allaire2002five} was chosen as framework that can be used to simulate many different conditions in multiphase flows. The present formulation slightly differs from the original model by considering an additional equation for an advected function $\psi$, which is used to compute the interface normal vectors in a similar fashion with respect to the work by Al-Salami et al. \cite{al2021high}. It is worthwhile mentioning that the method proposed in \cite{al2021high} considered a weakly compressible formulation. In this work, the same approach was generalised to the fully compressible five equation model. Furthermore, their choice of parameters in the conservative phase field equation did not satisfy boundedness of the phase field variable, leading to non-negligible mass conservation errors. As it will be shown in the numerical results section, these errors are significantly smaller with the present formulation. 

The full system, including contributions for considering viscous, gravitational and surface tension effects reads:
\begin{align}
\label{eq:5eq1}
 \frac{\partial \phi_{1}}{\partial t} + \textbf{u} \cdot \nabla \phi_{1} &= \nabla \cdot \textbf{a}_{1}, \\
 \frac{\partial \psi}{\partial t} + \textbf{u} \cdot \nabla \psi &= 0, \label{eq:5eq_finalPC2}\\
 \frac{\partial \rho_{1} \phi_{1}}{\partial t} + \nabla \cdot (\rho_{1} \phi_{1} \textbf{u}) &= \nabla \cdot \textbf{R}_{1}, \\
 \frac{\partial \rho_{2} \phi_{2}}{\partial t} + \nabla \cdot (\rho_{2} \phi_{2} \textbf{u}) &= \nabla \cdot \textbf{R}_{2}, \\
 \frac{\partial \rho \textbf{u}}{\partial t} + \nabla \cdot (\rho \textbf{u}  \otimes \textbf{u} + P \mathbb{I}) &= \nabla \cdot (\textbf{f} \otimes \textbf{u}) + \nabla \cdot \boldsymbol{\tau} + \sigma \kappa \widehat{\textbf{n}} \delta_{\Gamma} + \rho \textbf{g}, \\
 \frac{\partial \rho E}{\partial t} + \nabla \cdot ((\rho E + P) \textbf{u}) &= \nabla \cdot (\textbf{f} k) + \sum_{l=1}^{2} \nabla \cdot ( \rho_{l} H_{l} \textbf{a}_{l}) + \nabla \cdot (\boldsymbol{\tau} \cdot \textbf{u}) \\ 
& + \sigma \kappa \textbf{u} \cdot \widehat{\textbf{n}} \delta_{\Gamma}+ \rho \textbf{g} \cdot \textbf{u},
\label{eq:5eq2}
\end{align}
where $\phi_{l}$ is the phase field associated to the $l$-th phase, $\rho_{l}$ is the density of each phase, $\psi$ is the additional level-set function, $\rho \textbf{u}$ is the total momentum, $P$ is the pressure and $\rho E = \rho e + \frac{1}{2}\rho ||\textbf{u}||^{2}$ is the total energy. 

In addition,
\begin{equation}
\textbf{a}_{l} = \Gamma( \epsilon \nabla \phi_{l} - \phi_{l} (1- \phi_{l}) \widehat{\textbf{n}}_{l}), \quad \textbf{R}_{l} = \rho_{l}^{(0)} \textbf{a}_{l}, \quad \textbf{f} = \sum_{l=1}^{2} \textbf{R}_{l}, \quad k=\frac{1}{2} ||\textbf{u}||^{2},
\end{equation}
and $H_{l}$ is the specific enthalpy of the $l$-th phase. In terms of modeling of buoyancy, viscous stresses and surface tension: $\boldsymbol{\tau}= 2 \mu (\mathbb{S} - 1/3(\nabla \cdot \textbf{u}) \mathbb{I})$ is the viscous stress tensor, with $\mu$ the dynamic viscosity of the mixture evaluated as $\mu = \mu_{1} \phi_{1} + \mu_{2} \phi_{2}$, $\mathbb{S} = (\nabla \textbf{u} + \nabla \textbf{u}^{\intercal})/2$ is the strain-rate tensor, $\textbf{g}$ is the gravitational acceleration, $\sigma$ is the surface tension coefficient, $\kappa = -\nabla \cdot \widehat{\textbf{n}}$ is the curvature of the interface and $\delta_{\Gamma} = ||\nabla \phi_{1}||$ is an approximate delta function around the interface.

The system is then closed by relating internal energy with the pressure field using an Equation of State (EOS). A classical choice is the Stiffened-Gas EOS \cite{harlow1971fluid}:
\begin{equation}
P = \frac{\rho e - \bigg ( \frac{ \gamma_{1} P^{\infty}_{1}}{\gamma_{1}-1} \phi_{1} + \frac{ \gamma_{2} P^{\infty}_{2}}{\gamma_{2}-1} \phi_{2} \bigg)}{ \bigg ( \frac{\phi_{1}}{\gamma_{1}-1} +   \frac{\phi_{2}}{\gamma_{2}-1} \bigg)},
\end{equation}
where $\gamma_{l}$ and $P^{\infty}_{l}$ are the parameters of the EOS. 
From the stiffened-gas equation of state it is possible to write the speed of sound and specific enthalpy of each phases as
\begin{equation}
c_{l} = \sqrt{\gamma_{l}\bigg( \frac{P + P^{\infty}_{l}}{\rho_{l}} \bigg)} \quad \mathrm{and} \quad H_{l} = \frac{(P + P^{\infty}_{l})\gamma_{l}}{\rho_{l} (\gamma_{l} -1)} \quad \mathrm{for} \quad l =1,2.
\end{equation}
Finally, for completeness, the following mixture relations apply:
\begin{align}
\phi_{2} = & 1- \phi_{1},   \\
\rho = & \rho_{1}\phi_{1} + \rho_{2} \phi_{2},  \\
\frac{1}{\gamma-1} = & \phi_{1} \frac{1}{\gamma_{1}-1} + \phi_{2} \frac{1}{\gamma_{2}-1}, \\
P^{\infty} \frac{\gamma}{\gamma-1} =  & \phi_{1} \frac{\gamma_{1} P_{1}^{\infty}}{\gamma_{1}-1} + \phi_{2} \frac{\gamma_{2} P_{2}^{\infty}}{\gamma_{2}-1}.
\end{align}
In order to preserve the \emph{Interface Equilibrium Condition} (IEC), the quasi-conservative formulation proposed by Cheng et al. \cite{cheng2020quasi} is herein employed.  In \cite{cheng2020quasi}, a DG discretisation was coupled with an explicit limiting technique to avoid oscillations of the phase field in proximity of the interface. In the present work, instead, the sharpening/diffusion balance first proposed by Chiu \& Lin \cite{chiu2011conservative} was used. Even though similar results can be achieved with both approaches, the simplicity of implementation is surely one of the main advantages of the present strategy. The same technique, in fact, can be adapted to any numerical scheme for appropriate choices of parameters and numerical fluxes.

The present formulation was implemented within the opensource code \emph{Quail} \cite{ching2022quail}.
\section{Discontinuous Galerkin discretisation}\label{sec:DG}
In this section the Discontinuous Galerkin (DG) method will be briefly introduced by considering a general system of conservation laws. The same strategy is then applied to the discretisation of the  phase field equation and subsequently to the compressible five equation model for two-phase flows.

A general set of conservation laws can be written in a compact form as
\begin{equation}
\pd{\textbf{w}}{t} + \nabla \cdot \textbf{F}(\textbf{w},\nabla \textbf{w}) = \textbf{S}(\textbf{w}, \nabla \textbf{w}).
\label{EQN_massmomeng2}
\end{equation}
Before applying the DG approach to equation \ref{EQN_massmomeng2}, we first introduce the computational domain $\Omega$ as the partition of $N_{e}$ non-overlapping discrete
elements such that $\Omega= \cup_{n=1}^{N_{e}} \Omega_{n}$. Let us also denote the boundary of the $n$-th element as $\partial \Omega_{n}$. 

Secondly, we introduce the space of test functions in which the numerical solution will be seeked into. In particular, a classical choice for DG scheme consist in the functional space:
\begin{equation}
\mathcal{V}:= \{ \varphi \in L^{2}(\Omega): \varphi \arrowvert_{\Omega_{n}} \in \mathbb{P}_{\mathrm{p}}(\Omega_{n}), \forall \Omega_{n} \},
\end{equation}
where $\mathbb{P}_{\mathrm{p}}$ is the space of piecewise continuous polynomials of order not greater than p on $\Omega_{n}$ and 
\begin{equation}
L^{2} (\Omega) = \bigg \{ \varphi : \Omega_{n} \rightarrow \mathbb{R} \bigg \arrowvert \int_{\Omega_{n}}|\varphi(\textbf{x})|^{2} d \Omega \leq \infty \bigg \}
\end{equation}
with $\mathbb{R}$ being the space of the real numbers. Different bases can be used to define the functional space $\mathcal{V}$, both modal or nodal \cite{hesthaven:book}. In this work, classical Lagrange polynomials are used within a modal DG framework.

The approximation of the global solution $\textbf{w}^{\delta}$ can be defined as
\begin{equation}
\textbf{w}^{\delta} = \oplus_{n=1}^{N_{e}} \textbf{w}^{\delta}_{n},
\end{equation}
where $\textbf{w}^{\delta}_{n}$ is the local discrete solution: 
\begin{equation}
\textbf{w}^{\delta}_{n} = \sum_{j=0}^{\mathrm{p}} \textbf{w}_{j}(t) \varphi_{j}(\textbf{x}).
\label{eq:w_delta}
\end{equation}
The local formulation of the DG method requires $\textbf{w}^{\delta}_{n}$ to satisfy 
\begin{equation}
\int_{\Omega_{n}} \varphi_{i} \pd{\textbf{w}^{\delta}_{n}}{t} d\Omega + \int_{\Omega_{n}} \varphi_{i} \nabla \cdot \textbf{F}(\textbf{w}^{\delta}_{n},\nabla \textbf{w}^{\delta}_{n}) d\Omega = \int_{\Omega_{n}} \varphi_{i}  \textbf{S}(\textbf{w}_{n}^{\delta}, \nabla \textbf{w}_{n}^{\delta}) d\Omega \quad \forall \varphi_{i} \in \mathcal{V}.
\label{weakform}
\end{equation}
The second term on the left-hand side of equation \ref{weakform} is the flux term. Upon performing integration by parts, this term can be expressed as
\begin{equation}
\int_{\Omega_{n}} \varphi_{i} \nabla \cdot \textbf{F}(\textbf{w}^{\delta}_{n},\nabla \textbf{w}^{\delta}_{n}) d\Omega  = - \int_{\Omega_{n}}  \nabla \varphi_{i} \cdot \textbf{F}(\textbf{w}^{\delta}_{n},\nabla \textbf{w}^{\delta}_{n}) d\Omega 
 + \oint_{\partial \Omega_{n}} \varphi_{i}\widehat{\textbf{F}}(\textbf{w}^{\delta, +}_{n}, \textbf{w}^{\delta, -}_{n},\nabla \textbf{w}^{\delta,+}_{n},\nabla \textbf{w}^{\delta,-}_{n},\widehat{\textbf{m}})dS,
\end{equation}
where $\widehat{\textbf{m}}$ is the outward-pointing unit normal vector, $(\cdot)^{+}$ and $(\cdot)^{-}$ denote the right and left state with respect to the element's interface $\partial \Omega_{n}$ and $\widehat{\textbf{F}}$ is the numerical flux.

After exploiting the form of $\textbf{w}^{\delta}_{n}$ (equation \ref{eq:w_delta}), the local discrete weak formulation reads:
\begin{equation}
\sum_{j=0}^{\mathrm{p}} \der{ \textbf{w}_{j}}{\mathrm{t}} \mathrm{M}_{ij}  = \int_{\Omega_{n}}  \nabla \varphi_{i} \cdot \textbf{F}(\textbf{w}^{\delta}_{n},\nabla \textbf{w}^{\delta}_{n}) d\Omega - \oint_{\partial \Omega_{n}} \varphi_{i} \widehat{\textbf{F}}(\textbf{w}^{\delta, +}_{n}, \textbf{w}^{\delta, -}_{n},\nabla \textbf{w}^{\delta,+}_{n},\nabla \textbf{w}^{\delta,-}_{n},\widehat{\textbf{m}})dS +  \int_{\Omega_{n}} \varphi_{i}  \textbf{S}(\textbf{w}_{n}^{\delta}, \nabla \textbf{w}_{n}^{\delta}, \textbf{x}) d\Omega,
\label{weakDG}
\end{equation}
where $\mathrm{M}_{ij} = \int_{\Omega_{n}} \varphi_{i} \varphi_{j} d \Omega$ represents the $(i,j)$-th entry of the element-local mass matrix.
Both volume and surface integrals appearing in equation \ref{weakDG} can be either computed analytically or using appropriate quadrature rules. Equation \ref{weakDG} can then be discretised in time with explicit or implicit schemes.

The only component of the spatial discretisation which is intrinsically dependent on the specific conservation law is the definition of the numerical fluxes, which needs to take into account the eigen-structure of the hyperbolic system. Lax-Friedrichs and Symmetric Interior Penalty \cite{hartmann2008optimal} fluxes will be considered as they offer good flexibility for a wide range of different conservation laws. 

Now that the general framework of the DG method is introduced, we will proceed in presenting the specific numerical strategy used in the discretisation of the five equation model.
First, the numerical treatment of the phase field and level-set equations will be considered in order to highlight the key aspects of the interface capturing technique only. The subsequent coupling with the five equation model will naturally follow with appropriate choices of numerical fluxes.

The first two equations of the five equation model introduced in the previous section read:
\begin{align}
\frac{\partial \phi_{1}}{\partial t} + \textbf{u} \cdot \nabla \phi_{1} &= \nabla \cdot \textbf{a}_{1},  \\
\frac{\partial \psi}{\partial t} + \textbf{u} \cdot \nabla \psi &= 0,
\end{align}
with $\textbf{a}_{1} = \Gamma( \epsilon \nabla \phi_{1} - \phi_{1} (1- \phi_{1}) \widehat{\textbf{n}}_{1})$ and $\widehat{\textbf{n}}_{1} = \nabla \psi/||\nabla \psi||$.

For a prescribed velocity field, this system is well-posed and it can be numerically resolved as it is common practice in kinematic tests for the validation of interface capturing techniques. In the following discussion, in the phase field equation, the subscript of $\phi_{1}$ will be dropped since only one phase field needs to be resolved (the phase field for the secondary fluid will be directly evaluated as $\phi_{2}=1-\phi_{1}$).

The local weak formulation of the classic Discontinuous Galerkin method for this set of equations reads:
\begin{align}
\int_{\Omega_{n}} \frac{\partial \phi_{n}}{\partial t} \varphi_{i} d \Omega+ \int_{\Omega_{n}} (\textbf{u} \cdot \nabla \phi_{n}) \varphi_{i} d \Omega& = -\int_{\Omega_{n}} \nabla \varphi_{i} \cdot \textbf{a}_{l} d\Omega +\oint_{\partial \Omega_{n}} \widehat{\textbf{a}}_{l} \cdot \widehat{\textbf{m}} \varphi_{i} d S,  \\
\int_{\Omega_{n}} \frac{\partial \psi_{n}}{\partial t} \varphi_{i} d \Omega+ \int_{\Omega_{n}} (\textbf{u} \cdot \nabla \psi_{n}) \varphi_{i} d \Omega& =0,
\end{align}
where the vector $\widehat{\textbf{a}}_{l}$ is the numerical flux at the interface between neighbouring elements which depends on the left and right state of the conservative variables and their gradients (namely, $\phi^{\pm}$, $\psi^{\pm}$, $\nabla \phi^{\pm}$, $\nabla \psi^{\pm}$). A symmetric interior penalty flux was chosen in the present work to evaluate this term. The transport term, instead, is treated as a source term. The system written as it is can be numerically resolved for a prescribed, analytical velocity field. Instead, when considering the fully coupled system where the velocity field is itself an unknown of the problem and modelling the advection term as a source term is needed in order to fulfil the interface equilibrium condition (\ie exact resolution of contact discontinuities). This property is of fundamental importance for compressible flows in order to avoid pressure and velocity oscillations in proximity of the interface. These oscillations can ultimately lead to global instability of the numerical scheme and thus the IEC represents a key aspect in the success of the simulation.

Integrating by parts twice, the previous system can be written as:
\[
\int_{\Omega_{n}} \frac{\partial \phi_{n}}{\partial t} \varphi_{i} d \Omega+ \int_{\Omega_{n}} (\textbf{u} \cdot \nabla \phi_{n}) \varphi_{i} d \Omega- \int_{\partial \Omega_{n}} \widehat{\textbf{m}} \cdot \widetilde{(\textbf{u} \phi_{n})} \varphi_{i} dS + \int_{\partial \Omega_{n}} \widehat{\textbf{m}} \cdot \widehat{(\textbf{u} \phi_{n})} \varphi_{i} dS= 
\]
\[
= -\int_{\Omega_{n}} \nabla \varphi_{i} \cdot \textbf{a}_{l} d\Omega + \oint_{\partial \Omega_{n}} \widehat{\textbf{a}}_{l} \cdot \widehat{\textbf{m}} \varphi_{i} d S 
\]
\[
\int_{\Omega_{n}} \frac{\partial \psi_{n}}{\partial t} \varphi_{i} d \Omega+ \int_{\Omega_{n}} (\textbf{u} \cdot \nabla \psi_{n}) \varphi_{i} d \Omega- \int_{\partial \Omega_{n}} \widehat{\textbf{m}} \cdot \widetilde{(\textbf{u} \psi_{n})} \varphi_{i} dS + \int_{\partial \Omega_{n}} \widehat{\textbf{m}} \cdot \widehat{(\textbf{u} \psi_{n})} \varphi_{i} dS=0
\]
where two additional numerical fluxes for the transport term need to be defined  (namely, $\widehat{(\cdot)}$ and $\widetilde{(\cdot)}$). The same expressions proposed by \cite{cheng2020quasi} were considered in this work. In particular,
\begin{equation}
\widehat{\textbf{m}} \cdot \widetilde{(\textbf{u} \phi)} =
\begin{cases}
\widehat{\textbf{m}} \cdot \textbf{u}_{n}^{-} \phi^{-}_{n} \enspace  \mathrm{in} \enspace \Omega^{-}_{n} \\
\widehat{\textbf{m}} \cdot \textbf{u}_{n}^{+} \phi^{+}_{n} \enspace  \mathrm{in} \enspace \Omega^{+}_{n} \\
\end{cases}
\enspace  \mathrm{and} \enspace 
\widehat{\textbf{m}} \cdot \widehat{(\textbf{u} \phi)} =
\begin{cases}
\widehat{\textbf{m}} \cdot \textbf{u}^{-}\{\!\!\{ \phi \}\!\!\} - c [\![ \phi ]\!]   \enspace  \mathrm{in} \enspace \Omega^{-}_{n} \\
\widehat{\textbf{m}} \cdot \textbf{u}^{+}\{\!\!\{ \phi \}\!\!\} - c [\![ \phi ]\!]  \enspace  \mathrm{in} \enspace \Omega^{+}_{n} \\
\end{cases}
\end{equation}
where $c$ represents the maximum characteristic speed of the system (\ie the speed of sound for the five equation model),
\[
\{\!\!\{ \phi \}\!\!\} = \frac{\phi^{+}+\phi^{-}}{2} \quad \mathrm{and} \quad [\![ \phi ]\!] = \frac{\phi^{+}-\phi^{-}}{2}.
\]
It can be proven that this choice, in addition to the quasi-conservative form of the advection term, leads to an exact conservation of contact discontinuities if coupled with standard Lax-Friedrichs numerical flux for the remaining conservation laws in the five equation model (\ie conservation of mass  of the $l$-th phase, total momentum and total energy). For consistency, the same fluxes are employed also in the level-set equation even if this choice is not fundamental in the fulfilment of the interface equilibrium condition.

The local weak formulation for each equation of the full system can be written for

the Allen-Cahn equation as:
\begin{align}
\int_{\Omega_{n}} \frac{\partial \phi_{n}}{\partial t} \varphi_{i} d \Omega& + \int_{\Omega_{n}} (\textbf{u}_{n} \cdot \nabla \phi_{n}) \varphi_{i} d \Omega- \int_{\partial \Omega_{n}} \widehat{\textbf{m}} \cdot \widetilde{(\textbf{u}_{n} \phi_{n})} \varphi_{i} dS + \int_{\partial \Omega_{n}} \widehat{\textbf{m}} \cdot \widehat{(\textbf{u}_{n} \phi_{n})} \varphi_{i} dS=  \\
&= -\int_{\Omega_{n}} \nabla \varphi_{i} \cdot \textbf{a}_{1} d\Omega + \oint_{\partial \Omega_{n}} \widehat{\textbf{a}}_{1} \cdot \widehat{\textbf{m}} \varphi_{i} d S;
\end{align}

the level-set equation as:

\begin{align}
\int_{\Omega_{n}} \frac{\partial \psi_{n}}{\partial t} \varphi_{i} d \Omega& + \int_{\Omega_{n}} (\textbf{u}_{n} \cdot \nabla \psi_{n}) \varphi_{i} d \Omega- \int_{\partial \Omega_{n}} \widehat{\textbf{m}} \cdot \widetilde{(\textbf{u}_{n} \psi_{n})} \varphi_{i} dS + \int_{\partial \Omega_{n}} \widehat{\textbf{m}} \cdot \widehat{(\textbf{u}_{n} \psi_{n})} \varphi_{i} dS=0;
\end{align}

the mass conservation of phase $1$ as:

\begin{align}
\int_{\Omega_{n}} \frac{\partial (\rho_{1}\phi_{1})_{n}}{\partial t} \varphi_{i} d \Omega&- \int_{\Omega_{n}} (\rho_{1}\phi_{1})_{n}  \textbf{u}_{n} \cdot \nabla \varphi_{i} d \Omega+ \oint_{\partial \Omega_{n}}  \widehat{(\rho_{1}\phi_{1})_{n}  \textbf{u}_{n}} \cdot \widehat{\textbf{m}} \varphi_{i} dS = \\
&= -\int_{\Omega_{n}} \nabla \varphi_{i} \cdot \textbf{R}_{1} d\Omega +\oint_{\partial \Omega_{n}} \widehat{\textbf{R}}_{1} \cdot \widehat{\textbf{m}} \varphi_{i} d S;
\end{align}

the mass conservation of phase $2$ as:

\begin{align}
\int_{\Omega_{n}} \frac{\partial (\rho_{2}\phi_{2})_{n}}{\partial t} \varphi_{i} d \Omega& - \int_{\Omega_{n}} (\rho_{2}\phi_{2})_{n}  \textbf{u}_{n} \cdot \nabla \varphi_{i} d \Omega+ \oint_{\partial \Omega_{n}}  \widehat{(\rho_{2}\phi_{2})_{n}  \textbf{u}_{n}} \cdot \widehat{\textbf{m}} \varphi_{i} dS = \\
&= -\int_{\Omega_{n}} \nabla \varphi_{i} \cdot \textbf{R}_{2} d\Omega +\oint_{\partial \Omega_{n}} \widehat{\textbf{R}}_{2} \cdot \widehat{\textbf{m}} \varphi_{i} d S;
\end{align}

the total momentum conservation as:

\begin{align}
\int_{\Omega_{n}} \frac{\partial (\rho \textbf{u})_{n}}{\partial t} \varphi_{i} d \Omega& - \int_{\Omega_{n}}  (\rho_{n} \textbf{u}_{n} \otimes \textbf{u}_{n} + P_{n} \mathbb{I}) \cdot \nabla \varphi_{i} d \Omega+ \oint_{\partial \Omega_{n}}  \widehat{(\rho_{n} \textbf{u}_{n} \otimes \textbf{u}_{n} + P_{n} \mathbb{I})} \cdot \widehat{\textbf{m}} \varphi_{i} dS = \\
&= -\int_{\Omega_{n}} \nabla \varphi_{i} \cdot (\textbf{f} \otimes \textbf{u}_{n}) d\Omega +\oint_{\partial \Omega_{n}} \widehat{(\textbf{f} \otimes \textbf{u}_{n})} \cdot \widehat{\textbf{m}} \varphi_{i} d S;
\end{align}

and the total energy conservation as:

\begin{align}
\int_{\Omega_{n}} \frac{\partial (\rho E)_{n}}{\partial t} \varphi_{i} d \Omega& - \int_{\Omega_{n}}  (\rho_{n}E_{n} + P_{n})\textbf{u}_{n} \cdot \nabla \varphi_{i} d \Omega+ \oint_{\partial \Omega_{n}}  \widehat{(\rho_{n}E_{n} + P_{n})\textbf{u}_{n} } \cdot \widehat{\textbf{m}} \varphi_{i} dS = \\
&= -\int_{\Omega_{n}} \nabla \varphi_{i} \cdot (\textbf{f} k + \sum_{l=1}^{2} \rho_{l}H_{l} \textbf{a}_{l}) d\Omega +\oint_{\partial \Omega_{n}} \widehat{(\textbf{f} k +\sum_{l=1}^{2} \rho_{l}H_{l} \textbf{a}_{l})} \cdot \widehat{\textbf{m}} \varphi_{i} d S.
\end{align}

The numerical fluxes related to the interface capturing technique (\ie terms involving $\textbf{a}_{l}$) are discretised using the symmetric interior penalty method whereas the remaining numerical convective fluxes are discretised using the Lax-Friedrichs flux.

Finally, it is relevant to present one additional detail related to the link between the phase field and the additional level-set function $\psi$ and their respective dynamics. The function $\psi$ can be interpreted as a smoother version of the phase field which is used to compute the interface normal vectors. 

However, it is easy to see that in the previous equations, for a given velocity field, the coupling is only one-way: the level-set equation will affect the phase field evolution due to the computation of the normal vectors but there is no information flowing from the phase field equation back to the dynamics of the level-set. Consequently, after sufficiently long time, the interface defined by the two functions will start to drift apart, leading to completely erroneous results.
Consequently, the function $\psi$ needs to be periodically re-initialised. In this work we followed the strategy proposed by \cite{al2021high} where every $2000$ time steps, the following re-initialisation equation
\begin{equation}
\frac{\partial \psi}{\partial \tau} + sgn(\psi_{0}) (1-||\nabla \psi||) = \nabla \cdot (\nu_{h} \nabla \psi),
\end{equation}
is solved in pseudo-time from the initial condition $\psi_{0} = \phi - \frac{1}{2}$. The quantity 
\[
sgn(\psi_{0}) = \tanh \bigg( \frac{\psi_{0}}{2\epsilon ||\nabla \psi_{0}||} \bigg)
\]
is a smeared sign function and $\nu_{h}$ is a vanishing viscosity proportional to the grid size. Since a smooth profile for the level-set function is only needed inside a narrow bandwidth localised in proximity of the interface, the re-initialisation equation is integrated only for a limited number of pseudo-time steps. It was found that applying approximately $50$ iterations every $2000$ physical time-steps was at the same time able to 
\begin{inparaenum}[(i)] 
\item provide a sufficiently smooth profile of the interface, 
\item avoid decoupling between phase field and level-set and
\item keep the computational cost of the re-initialisation process relatively low.
\end{inparaenum}
In similarity with the phase field equation, the numerical diffusive flux in the re-initialisation equation is discretised using the symmetric interior penalty approach whereas the smeared sign hyperbolic term is treated as a simple source term. It is worthwhile mentioning that even if a re-initialisation step is herein considered, the mass conservation errors associated to this process are identically equal to zero since the level-set function is only used to compute the interface normal vectors and not to track the interface (as it is normally done in classical level-set methods).

It is important to underline that the present methodology improves the previously cited works from several point of views, increasing either the accuracy, the simplicity of implementation or the generality of each model. The quasi-conservative approach presented in \cite{cheng2020quasi} is used to preserve contact discontinuities in the five equation model and it is herein further improved by avoiding explicit limiting of the volume fraction employing, instead, explicit sharpening/diffusive terms of the Allen-Cahn equation. In this way the proposed model is characterised by a significant simplicity of implementation and flexibility. For example, considering complex unstructured grids, the proposed methodology does not need any ad-hoc limiting which might become considerably expensive and complex to implement. The present approach also allows the use of much higher orders of approximation with respect to the ones used in \cite{cheng2020quasi}. The advantages of using high orders of approximation will be further discussed in the results section. In the same section, the proposed methodology will also be tested on triangular elements, highlighting the robustness of this approach for unstructured grids. Finally, the model herein presented extends the interface capturing technique proposed by \cite{al2021high} to a fully compressible formulation and it was also possible, by appropriate choices of the parameters, to significantly reduce mass conservation errors of the original technique.
\section{Results}\label{sec:results}
\subsection{Kinematic tests}
Regarding the validation of the present methodology, the proposed numerical test cases will be sorted in kinematic tests and two-phase flow tests. For the former case, the velocity field is represented by a given analytical function which will be used in the phase field and level-set equations. In this way, it is possible to assess the performance of the interface capturing technique in a segregated fashion. The latter set of tests considers the full five equation system which is used to describe the flow of compressible two-phase flows.
\subsubsection{Linear-advection of a droplet} 
The first test case herein considered is a simple advection of a circular droplet with a prescribed interface thickness. The goal of this specific test is to verify that the additional diffusive/sharpening terms, for a given hyperbolic tangent initial condition, do not deteriorate the order of accuracy of the underlying numerical scheme. The hyperbolic tangent profile, in fact, is supposed to maintain its shape and simply be transported by the constant advection velocity. 

Consequently, a spherical droplet of radius $R=0.15$, represented by the phase field
\begin{equation}
\phi(\textbf{x},0) = \frac{1}{2}\bigg(1 + \tanh \bigg( \frac{r-R}{2\epsilon}\bigg) \bigg) \quad \mathrm{with} \quad r = ||\textbf{x}||,
\end{equation}
is advected by the constant advection velocity $\textbf{u} = (1,0)$ in a $[0,1]^{2}$ periodic square for a full period. The parameter $\epsilon$ is set to be a constant equal to the coarsest mesh resolution (\ie $\epsilon=0.05$). 

\begin{figure}[h!]
\centering
\includegraphics[width=.75\textwidth]{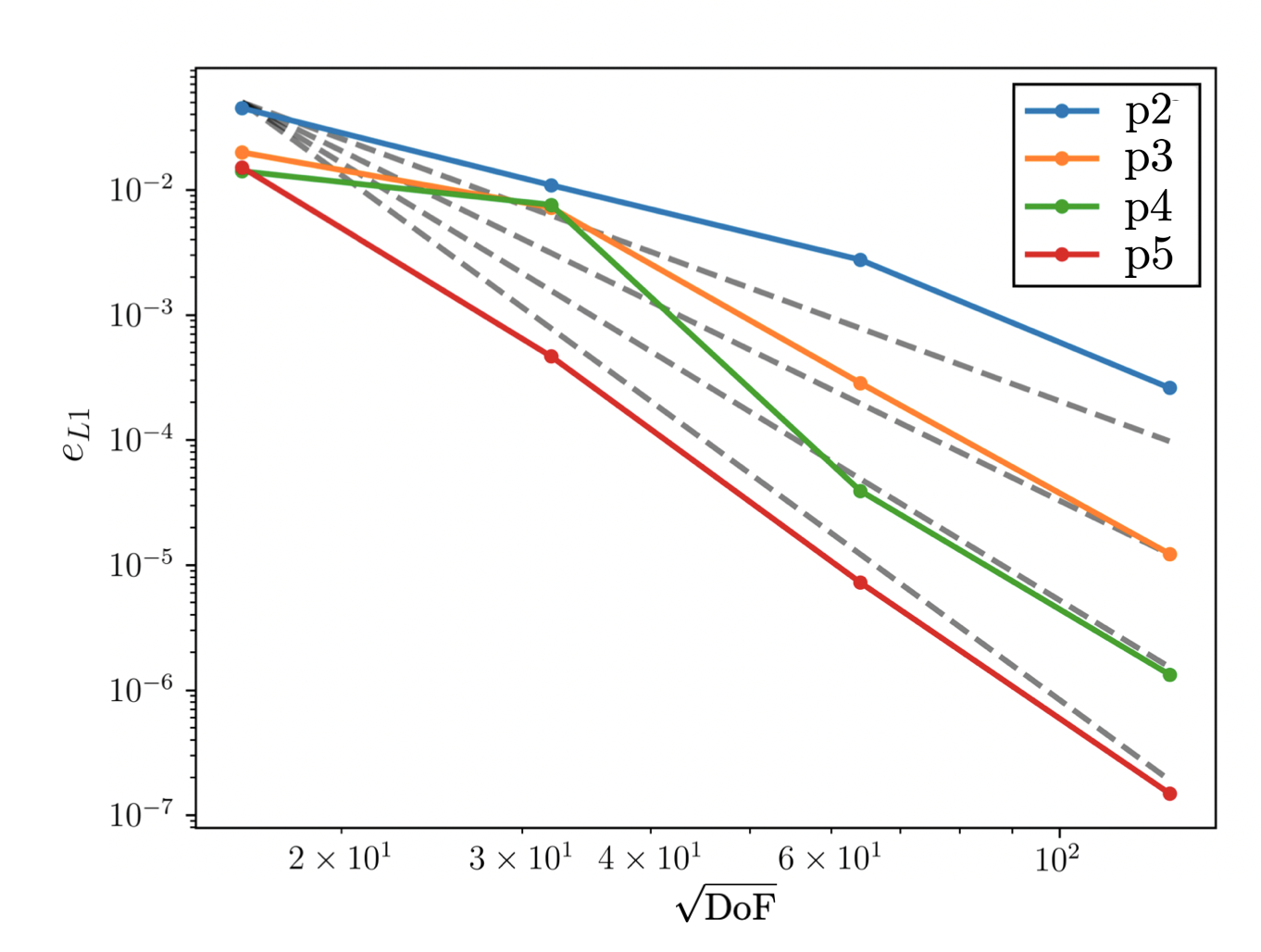}
\caption{$L^{1}$-error against number of elements along each direction for different orders of approximation. Optimal orders of convergence are shown by the dashed grey lines.}
\label{fig:conv_fix_eps}
\end{figure}

Similarly, the level-set, signed-distance function is initialised as:
\begin{equation}
\psi(\textbf{x},0) = -r.
\end{equation}
The $L^{1}$ error of the phase field as a function of the number of elements and different orders of approximation is shown in figure \ref{fig:conv_fix_eps}. It can be seen that the optimal order of convergence of the spatial discretisation is correctly recovered under mesh refinement.
%
\subsubsection{Rider-Kothe vortex}
Another standard test case for interface capturing techniques consists in the deformation of a circular bubble by a shear flow \cite{rider1998reconstructing}. In particular, a droplet of radius $R=0.15$ centered at $(0.0,0.25)$ in a $[-0.5,0.5]^{2}$ periodic domain is advected by the following divergence-free velocity field:
\begin{align}
u_{1} = & \sin^{2} (\pi x_{1}) \sin (2 \pi x_{2}) \cos \bigg( \frac{\pi t}{T}\bigg) \\
u_{2} = &-\sin ( 2 \pi x_{1}) \sin^{2} (\pi x_{2}) \cos \bigg( \frac{\pi t}{T}\bigg),
\end{align} 
where $T$ denotes the characteristic period of the shear flow. The classical value of $T=4$ is chosen for this particular test. Under the action of such a velocity field, in the first half of the period, the initial circular bubble is strongly deformed into a thin filament. After velocity reversal, the filament is stretched back to the initial condition. 

For this problem, we considered a polynomial discretisation of order $1$ and $3$ within the DG framework on a series of increasingly refined meshes. In particular, simulations involving $64^{2}$, $128^{2}$ and $256^{2}$ degrees of freedom were considered. Notice that in spectral element methods the total number of degrees of freedom is jointly defined by the number of elements and by the polynomial order of approximation. Consequently, for example, considering a $3^{\mathrm{rd}}$ polynomial order approximation on a $16\times16$ grid, the nomenclature will read: $16\times16$p$3$ (\ie $16$ elements times $16$ elements for a polynomial approximation of degree $3$). The total number of degrees of freedom is defined as $(16\times4)^{2}=64^{2}$. In comparisons of simulations involving different orders of approximation, a  common choice is to match the total number of degrees of freedom (for example, $16\times16$p$3$ and $32\times32$p$1$ simulations).

\begin{figure}[h!]
\centering
\subfigure[$16\times 16$ quadrilateral mesh (mesh $a$).]{\includegraphics[trim=40 0 40 0,clip,width=0.49\textwidth]{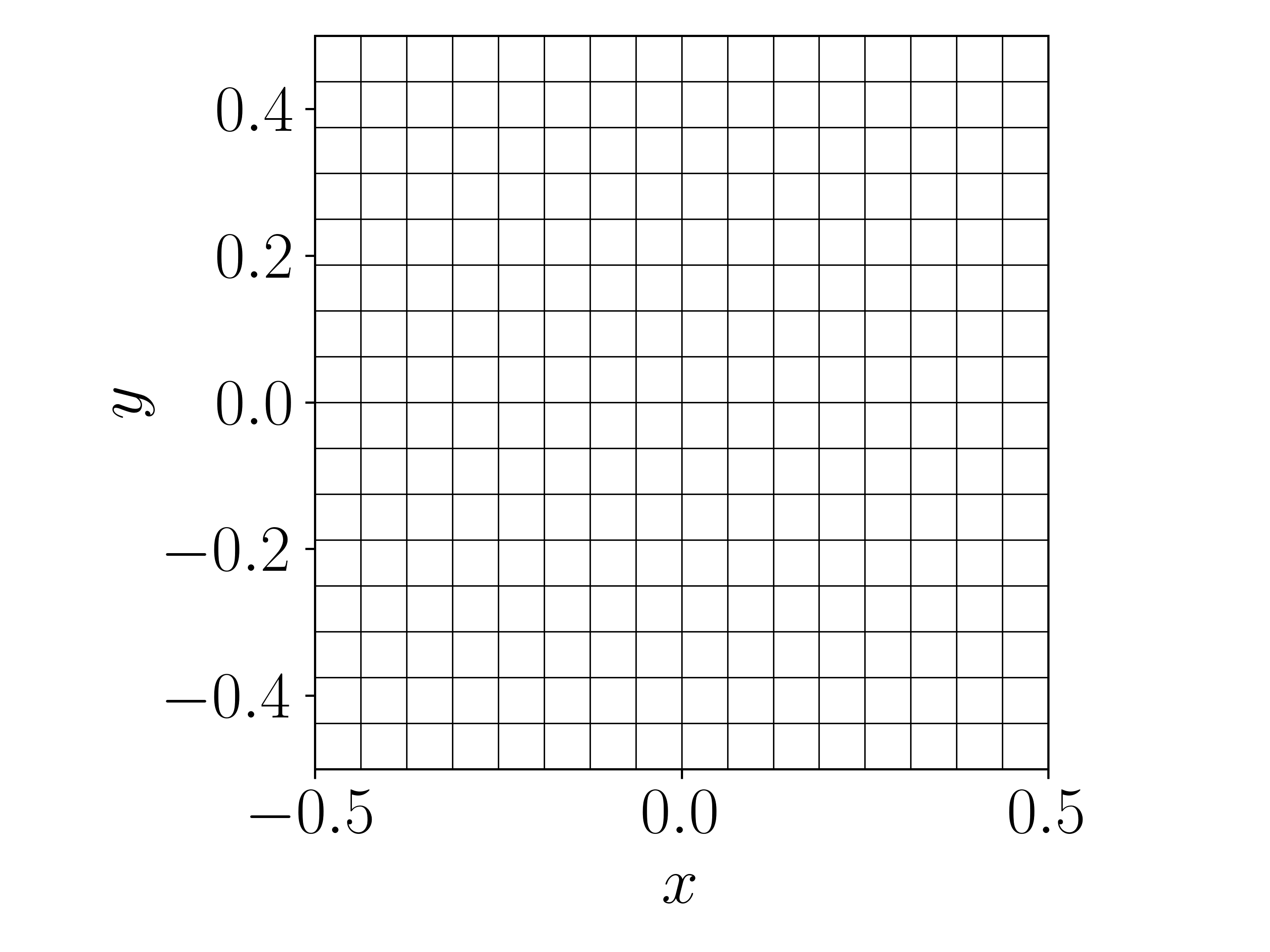}}
\subfigure[$16\times 16$ Triangular mesh (mesh $b$).]{\includegraphics[trim=40 0 40 0,clip,width=0.49\textwidth]{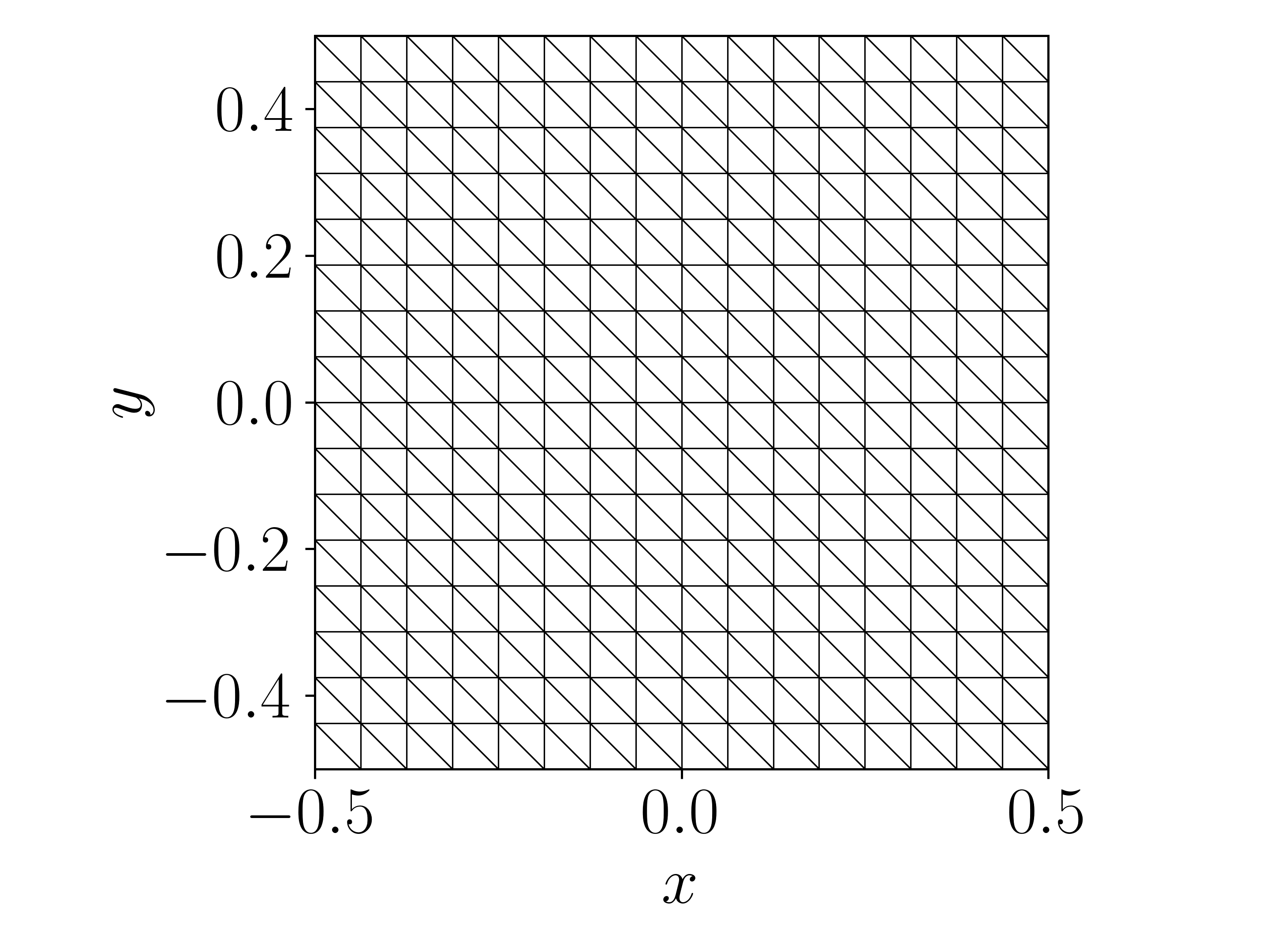}}
\caption{Examples of computational meshes used in the present work.}
\label{fig:meshes}
\end{figure}

\begin{figure}[h!]
\centering
\subfigure[Interface location at $t=T/2$.]{\includegraphics[trim=40 0 65 0,clip,width=0.49\textwidth]{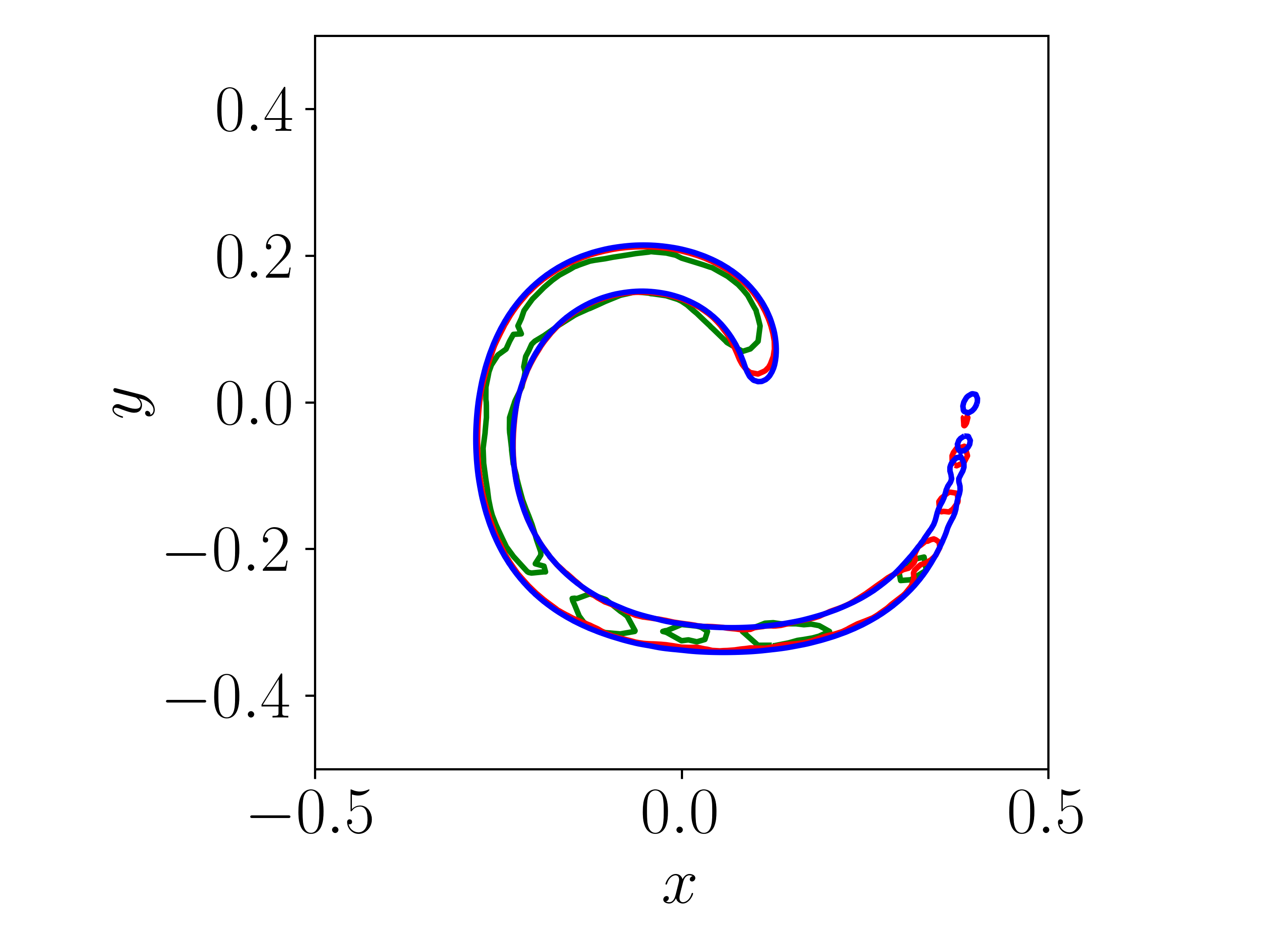}\label{fig:comparison_t1}}
\subfigure[Interface location at $t=T$.]{\includegraphics[trim=40 0 65 0,clip,width=0.49\textwidth]{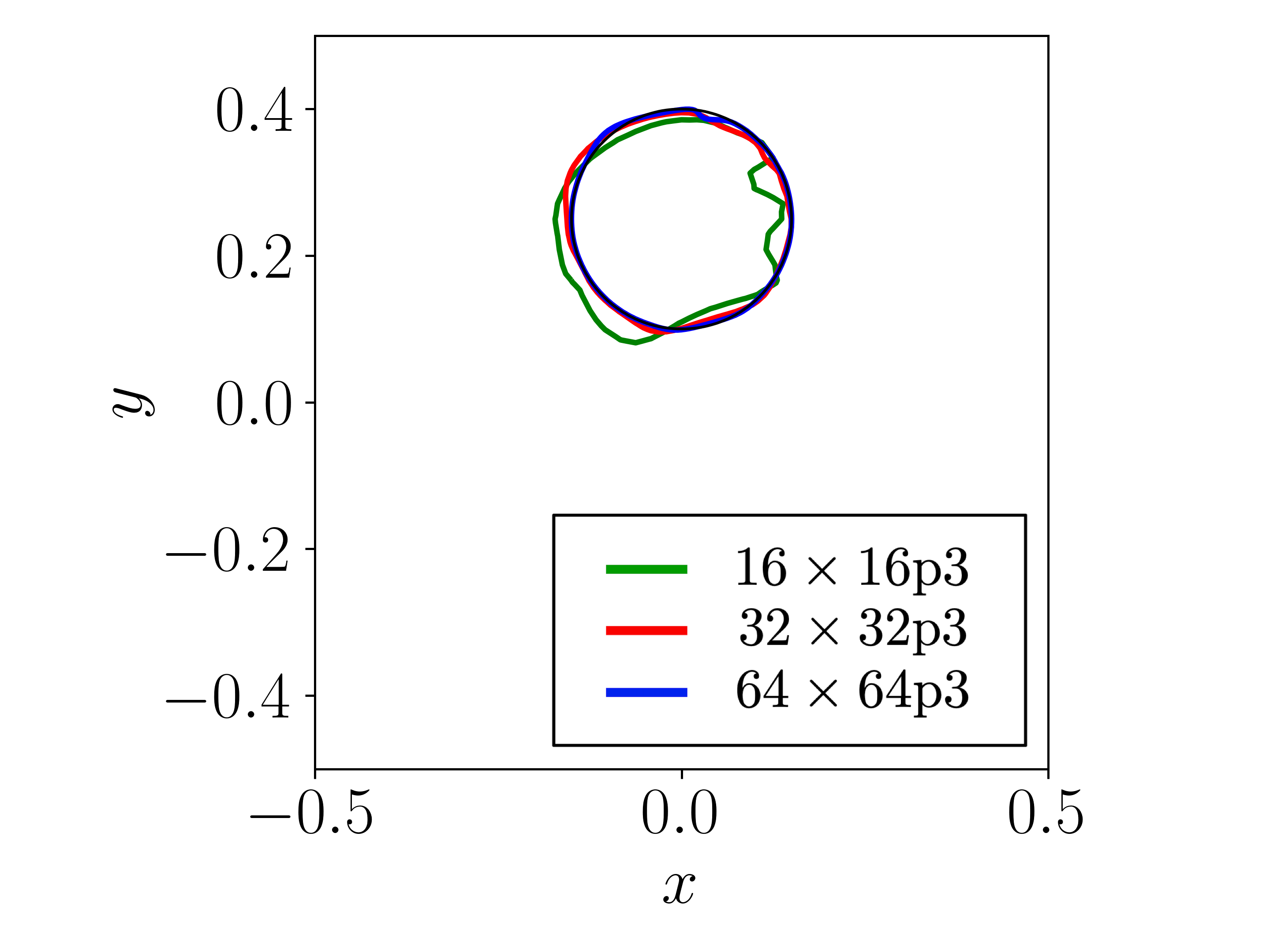}\label{fig:comparison_t2}}
\caption{Iso-contour $\phi =0.5$ after half period (left) and after one full period (right) for the $16\times16$p$3$, $32\times32$p$3$ and $64\times64$p$3$ simulations. The solid black line represents the initial location of the interface.}
\label{fig:comparison_t}
\end{figure}

\begin{figure}[h!]
\centering
\subfigure[Interface location at $t=T/2$.]{\includegraphics[trim=40 0 65 0,clip,width=0.49\textwidth]{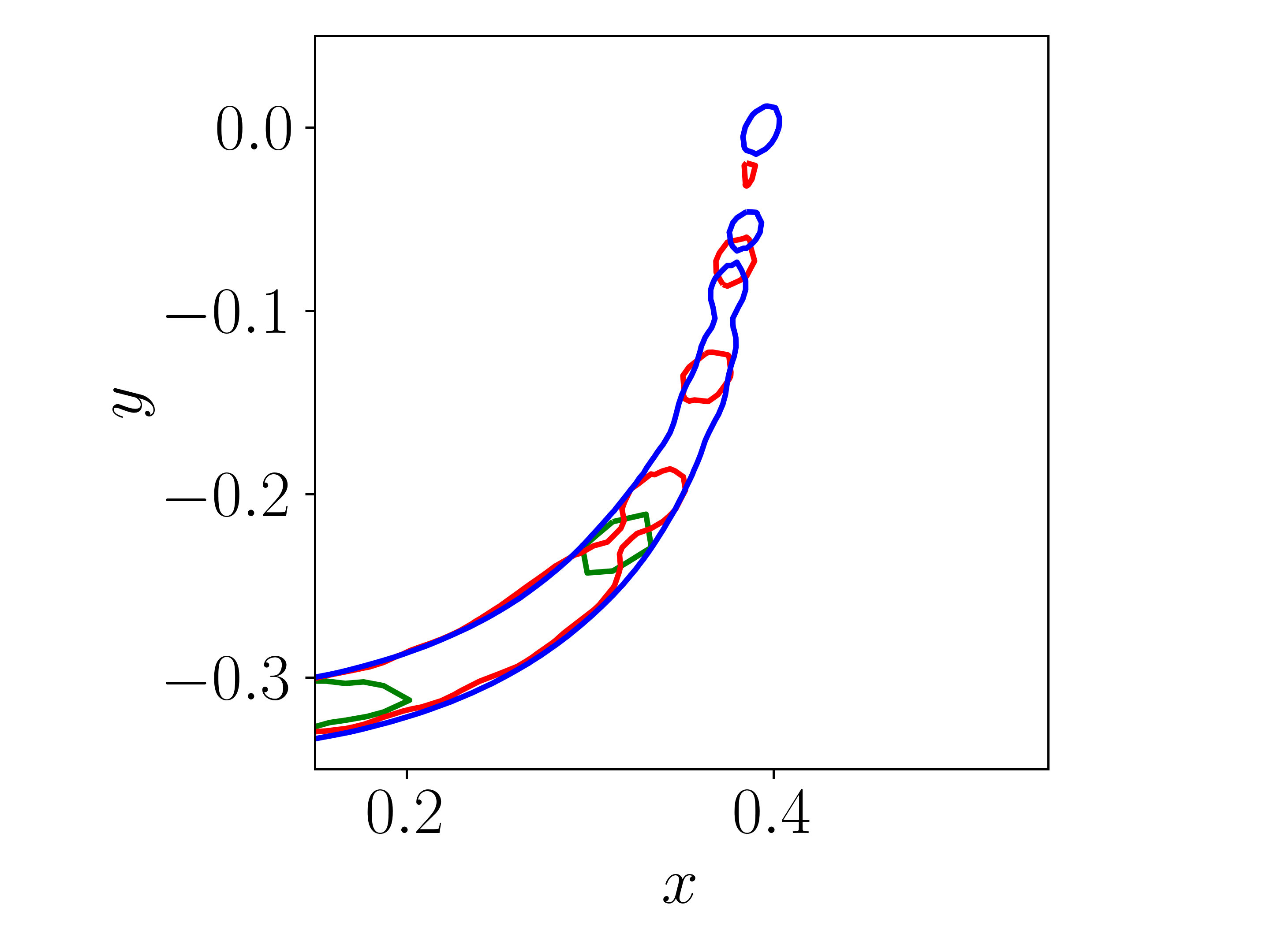}\label{fig:RK2}}
\subfigure[Interface location at $t=T$.]{\includegraphics[trim=40 0 65 0,clip,width=0.49\textwidth]{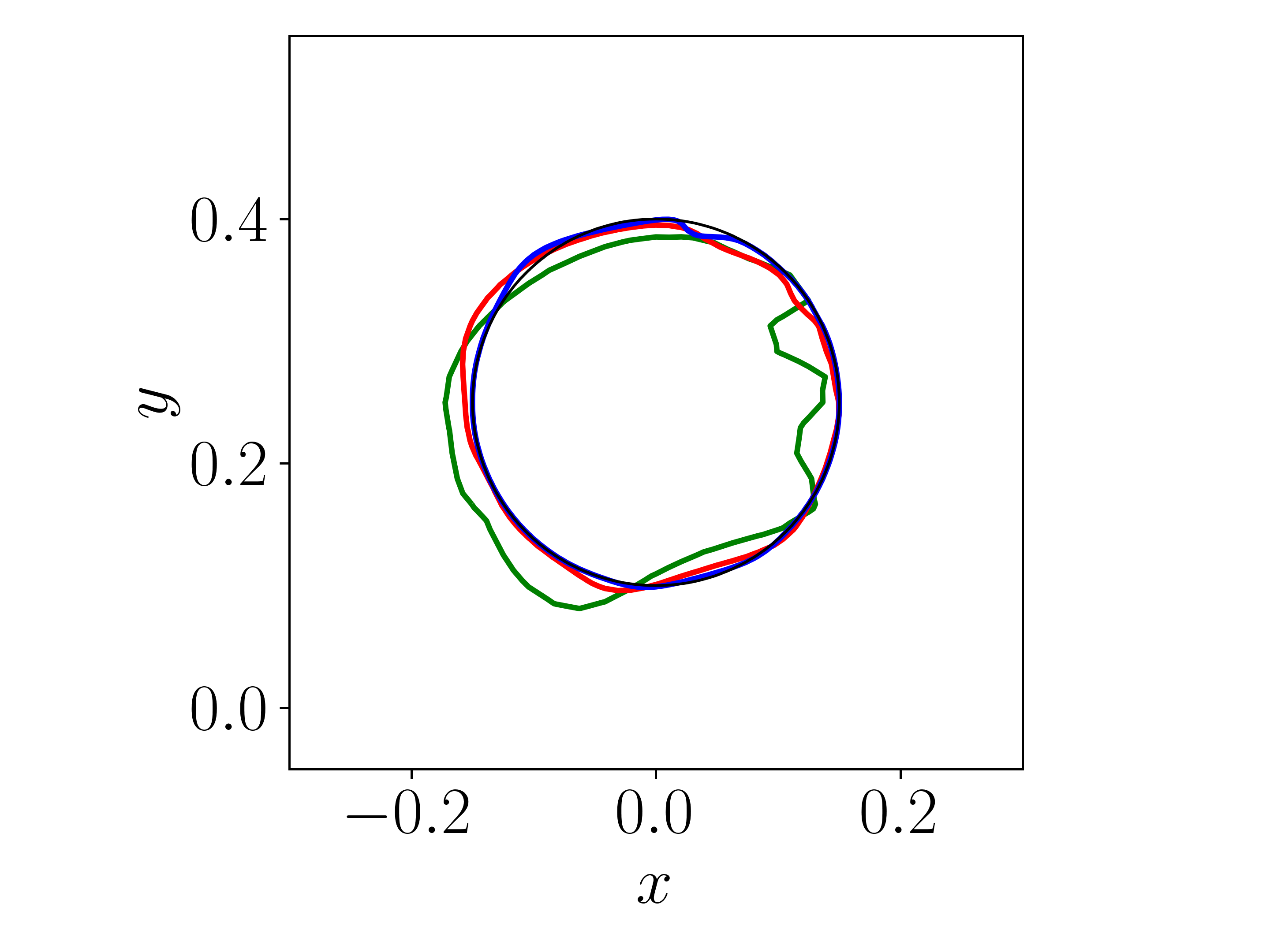}\label{fig:RK}}
\caption{Close-up view of the interface iso-contour at half period near breakup (left) and after one full period (right) for the $16\times16$p$3$, $32\times32$p$3$ and $64\times64$p$3$ simulations.}
\label{fig:detail}
\end{figure}

Time integration was performed using a classical $4^{\mathrm{th}}$ order Runge-Kutta scheme.
Also, in order to test the versatility of the proposed approach, both quadrilateral and triangular meshes were used. Examples of the meshes used in this work are shown in figure \ref{fig:meshes}.
In figure \ref{fig:comparison_t}, the interfacial profiles at half period and after a full period are shown for different levels of mesh refinement using a $3^{\mathrm{rd}}$ order polynomial approximation based on square elements. 
Two close-up views of figures \ref{fig:comparison_t1} and \ref{fig:comparison_t2} are shown in figure \ref{fig:detail} to better appreciate the convergence of the proposed approach. 

In figure \ref{fig:RK2}, we can observe that the resolution of the thin filament improves under mesh refinement. In figure \ref{fig:RK}, instead, the recovery of the initial condition can be better appreciated.  

\begin{figure}[h!]
\centering
\subfigure[$128^{2}$ DoF]{\includegraphics[trim=40 0 65 0,clip,width=0.49\textwidth]{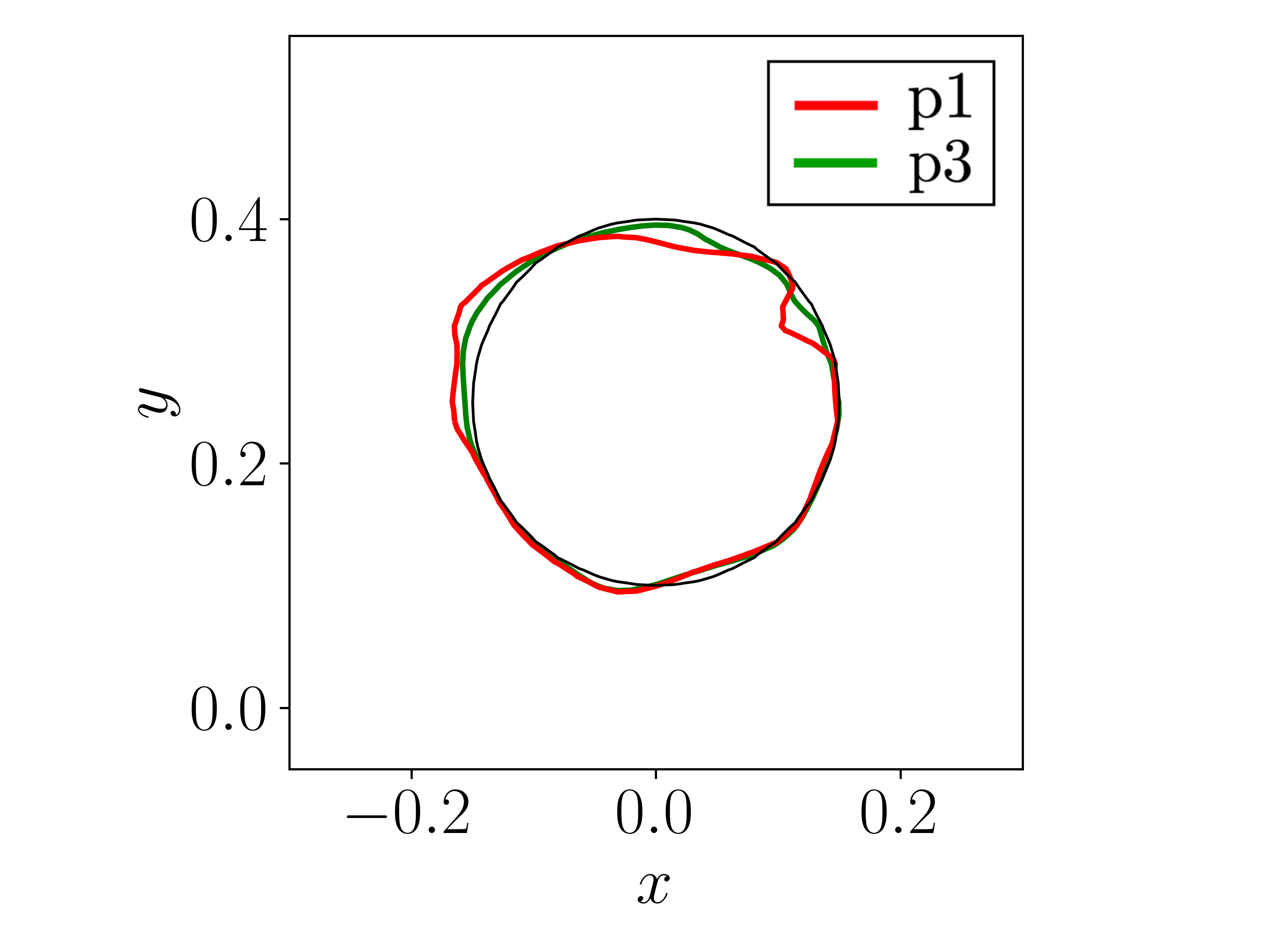}}
\subfigure[$256^{2}$ DoF]{\includegraphics[trim=40 0 65 0,clip,width=0.49\textwidth]{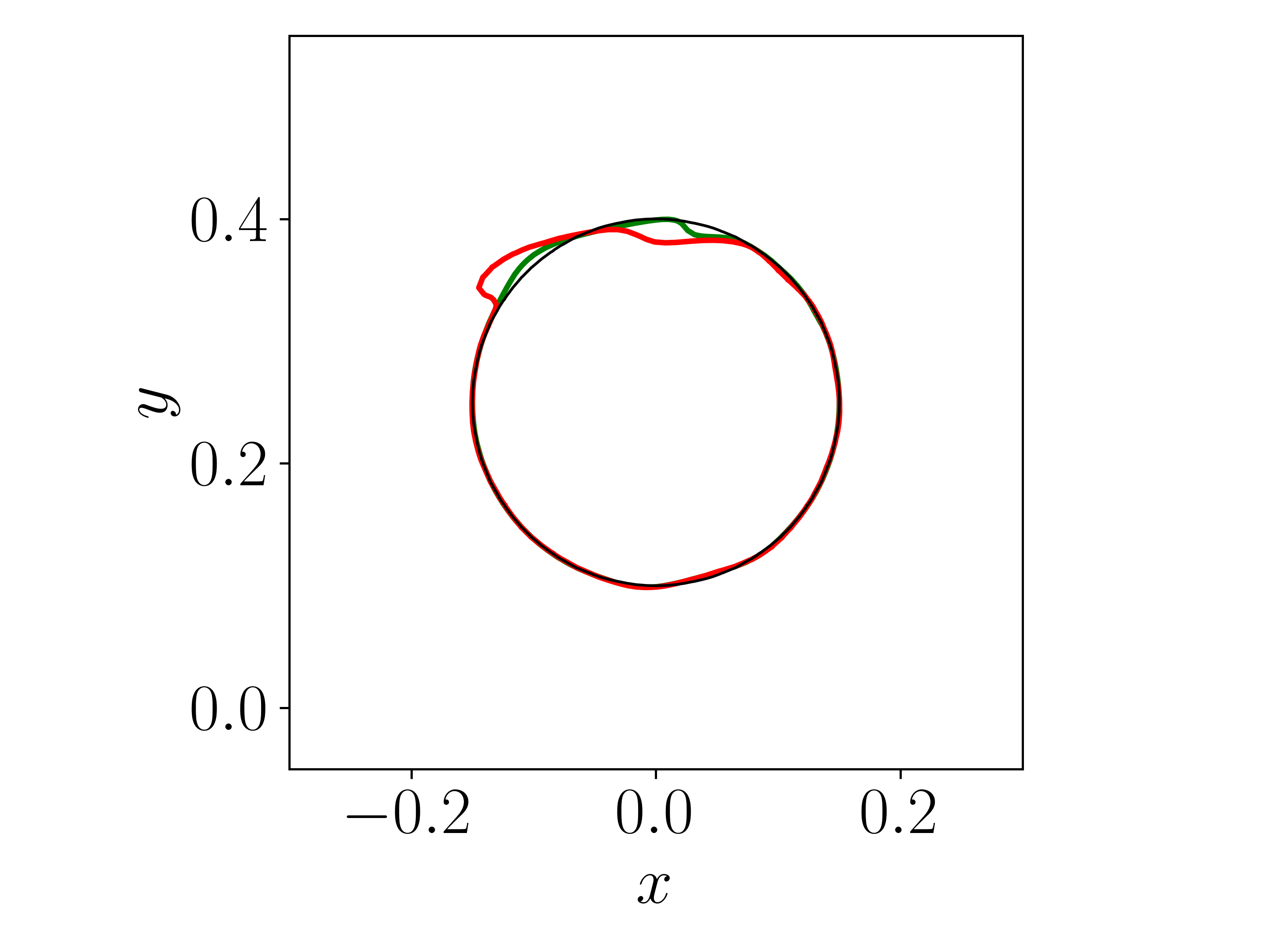}}
\caption{Iso-contour $\phi =0.5$ after one full period for $128^{2}$ (left) and $256^{2}$ DoF (right) with $1^{\mathrm{st}}$ and $3^{\mathrm{rd}}$ polynomial order approximations. The solid black line represents the initial location of the interface.}
\label{fig:comparison_p}
\end{figure}

In order to further investigate the proposed numerical strategy, lower order simulations using the same approach have been performed, keeping fixed the total number of degrees of freedom. In figure \ref{fig:comparison_p} the final location of the interface is shown for two different polynomial orders ($1$ and $3$) for two different resolutions ($128^{2}$ and $256^{2}$ DoF).

\begin{figure}[h!]
\centering
\includegraphics[width=.85\textwidth]{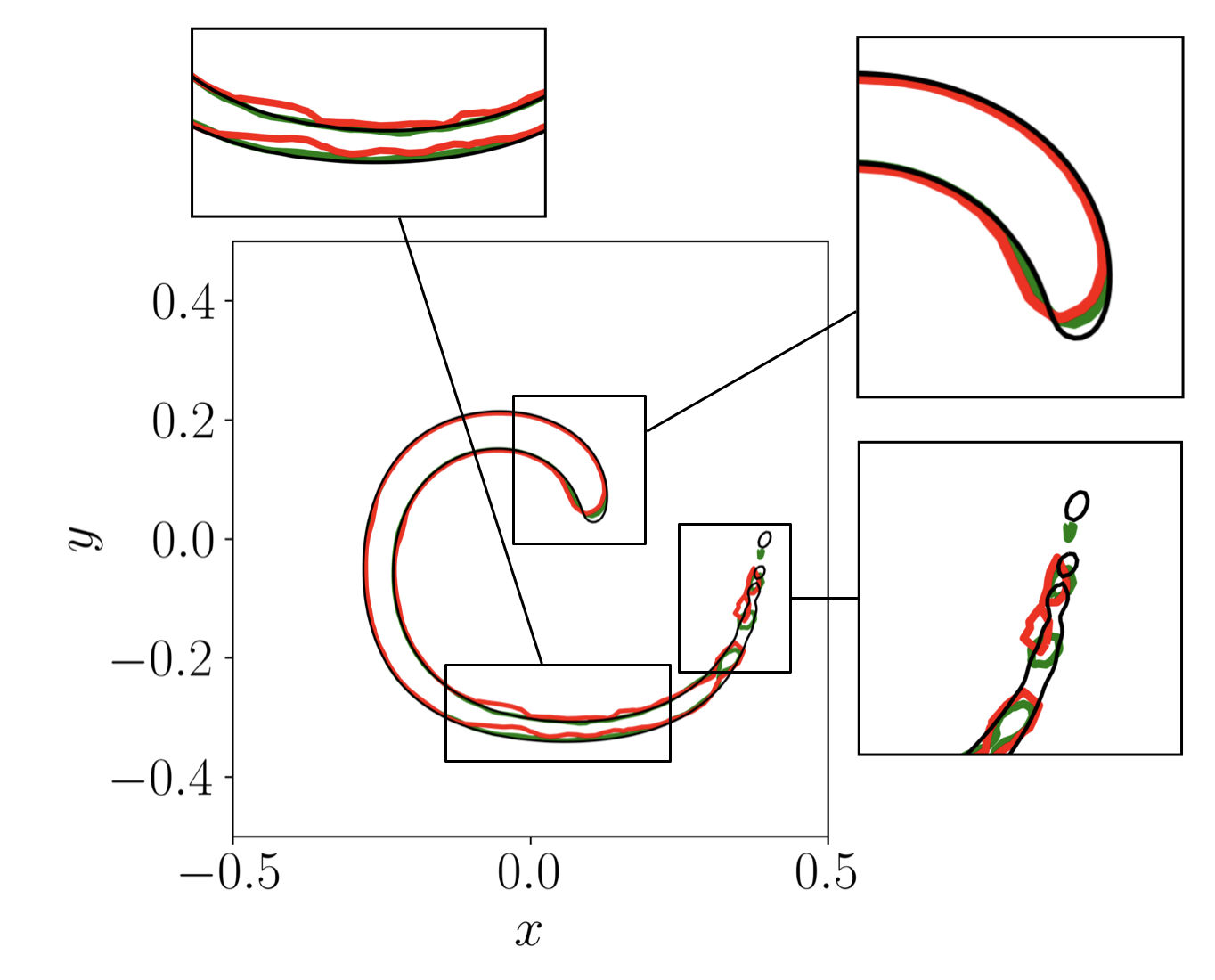}
\caption{Iso-contour $\phi =0.5$ after half period for $128^{2}$ DoF with $1^{\mathrm{st}}$  (\protect \redline) and $3^{\mathrm{rd}}$ (\protect \greenline) polynomial order approximations on rectangular elements. Some specific regions of the the interface are enlarge to highlight the difference between the two simulations. The simulation for $256^{2}$ DoF is used as reference.}
\label{fig:conv}
\end{figure}

It can be noticed that the higher order approximation provides more accurate results. For both resolutions, in fact, the p$3$ simulation outperforms the p$1$ computation in matching the exact initial solution. In particular, the upper part of the droplet is better approximated. In fact, this region is particularly stretched in the first half-period leading to significant under-resolution.
In order to further examine how under-resolution behaves, the location of the interface at the half-period is shown in figure \ref{fig:conv} for the two different polynomial orders. Different zones of the interface are highlighted to better appreciate the differences between the two simulations. It can be seen that in many locations the higher order approximation is generally smoother and it better matches the reference solution. In this case, the reference solution is simply a more refined simulation. 

These observations further highlight the significant improvement in considering high-order discretisations for the proposed phase field equation. In particular, it is the believe of the authors that low-order approximations not only introduce excessive amounts of numerical dissipation but also provide inaccurate evaluation of the normal vectors, therefore deteriorating the interface capturing technique. These assumptions will be further confirmed through additional numerical experiments for different orders of approximation. 
Secondly, in order to test the robustness and generality of the proposed approach, different types of finite elements were considered. In particular, in the following part of this section we compare quadrilateral and triangular elements (see figure \ref{fig:meshes}).

\begin{figure}[h!]
\centering
\subfigure[Interface location at $t=T/2$.]{\includegraphics[trim=40 0 65 00,clip,width=0.49\textwidth]{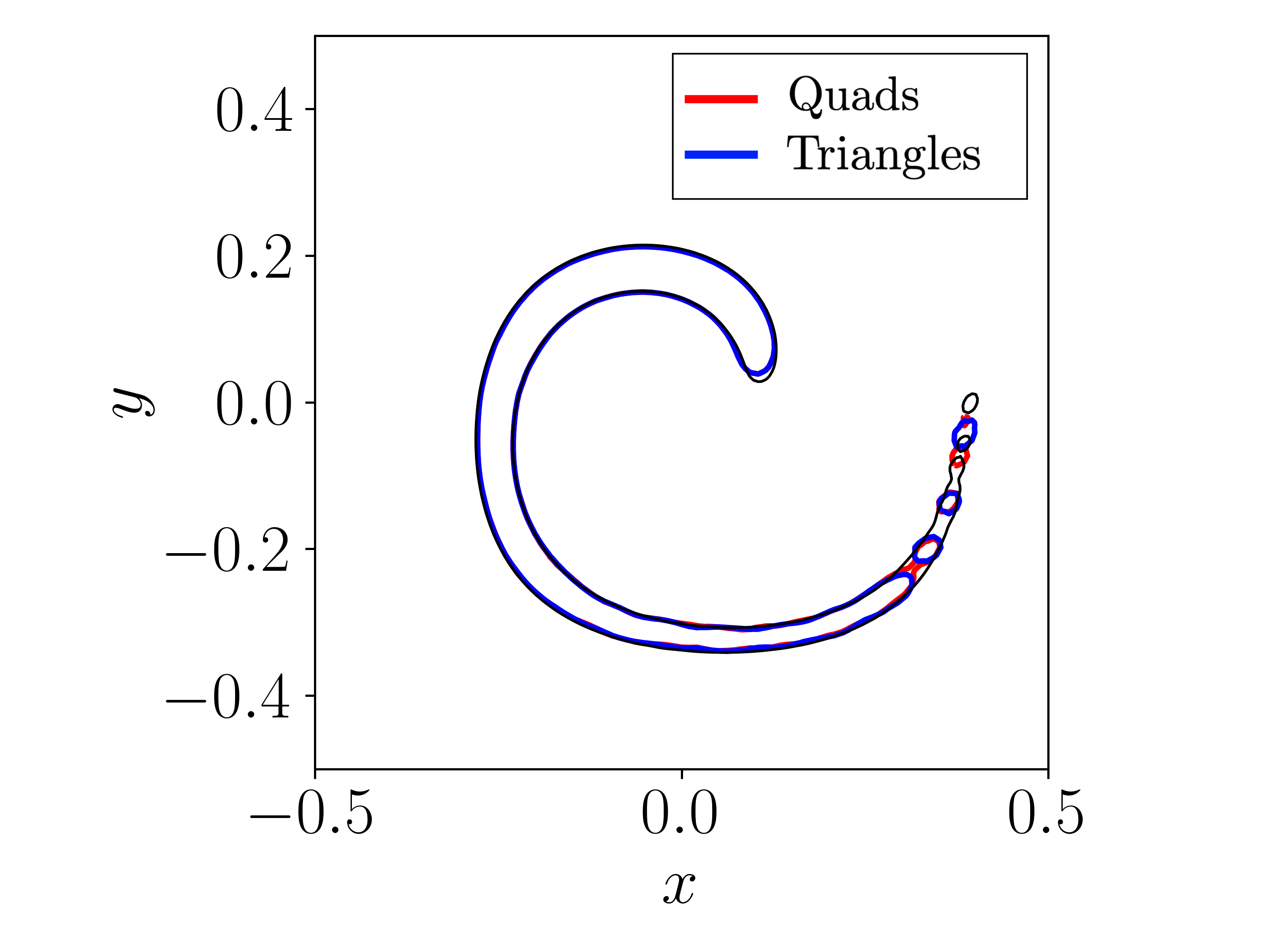}\label{fig:comparison_tr_half}}
\subfigure[Interface location at $t=T$.]{\includegraphics[trim=40 0 65 0,clip,width=0.49\textwidth]{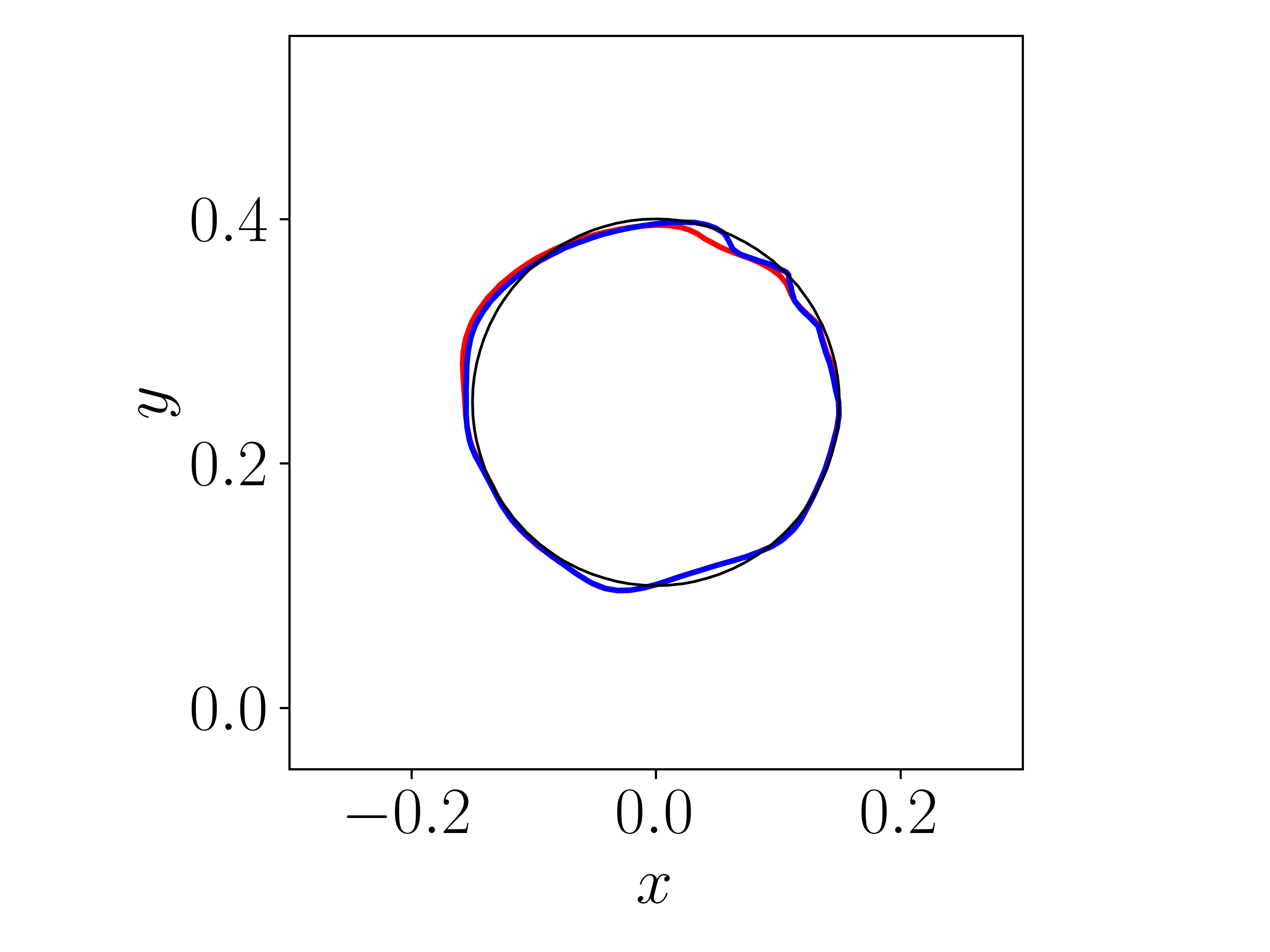}\label{fig:comparison_tr_full}}
\caption{Iso-contour $\phi =0.5$ after half period (left) and one full period (right) for the the $3^{\mathrm{rd}}$ polynomial order simulation with $128^{2}$ DoF comparing quadrilateral and triangular elements. The solid black line represents the reference solution (highly resolved simulation for $t=T/2$ and initial solution for $t=T$).}
\label{fig:comparison_tr}
\end{figure}

In figure \ref{fig:comparison_tr}, the location of the interface using quadrilateral and triangular elements is shown for $128^{2}$ DoF ($3^{\mathrm{rd}}$ polynomial order). In particular, the solution is shown at half-period on the left and after a full revolution on the right. 
In figure \ref{fig:comparison_tr_half} it is interesting to note that the two type of meshes (rectangular and triangular) provide almost exactly the same results in well-resolved regions of the domain, whereas noticeable differences can be observed only in proximity of the tail of the droplet. After one full revolution, the upper part of the droplet is slightly better approximated by triangular element as it can be observed in figure \ref{fig:comparison_tr_full}. 
It is important to highlight that such results further confirm the robustness and generality of the present approach. Using different types of elements did not require any ad-hoc modification of the underlying technique, which emphasises the simplicity and flexibility of the proposed implementation. The capability of handling triangular elements without deep modifications in the interface capturing techniques is a very desirable feature for the simulation of more complex configurations of engineering interest where unstructured meshes are often required.
In a more quantitative way, the $L^{1}$-errors were computed and are listed in table \ref{tab:L1RK} where they are compared with simulations from the literature involving the same number of degrees of freedom and similar interface capturing techniques (\ie based on the phase field approach). It can be observed that the present results are in good agreement with previously published simulations of the same test case. In the same table, it can also be noted that the differences between $1^{\mathrm{st}}$ and $3^{\mathrm{rd}}$ polynomial order approximations observed in previous plots, can now be quantified more clearly: the $L^{1}$-errors in the p$3$ simulations are always smaller than the p$1$ counterparts (sometimes even twice/three times smaller). Also, the gap between the two approximation orders tends to grow under mesh refinement.

\begin{figure}[h!]
\centering
\subfigure{\includegraphics[width=0.49\textwidth]{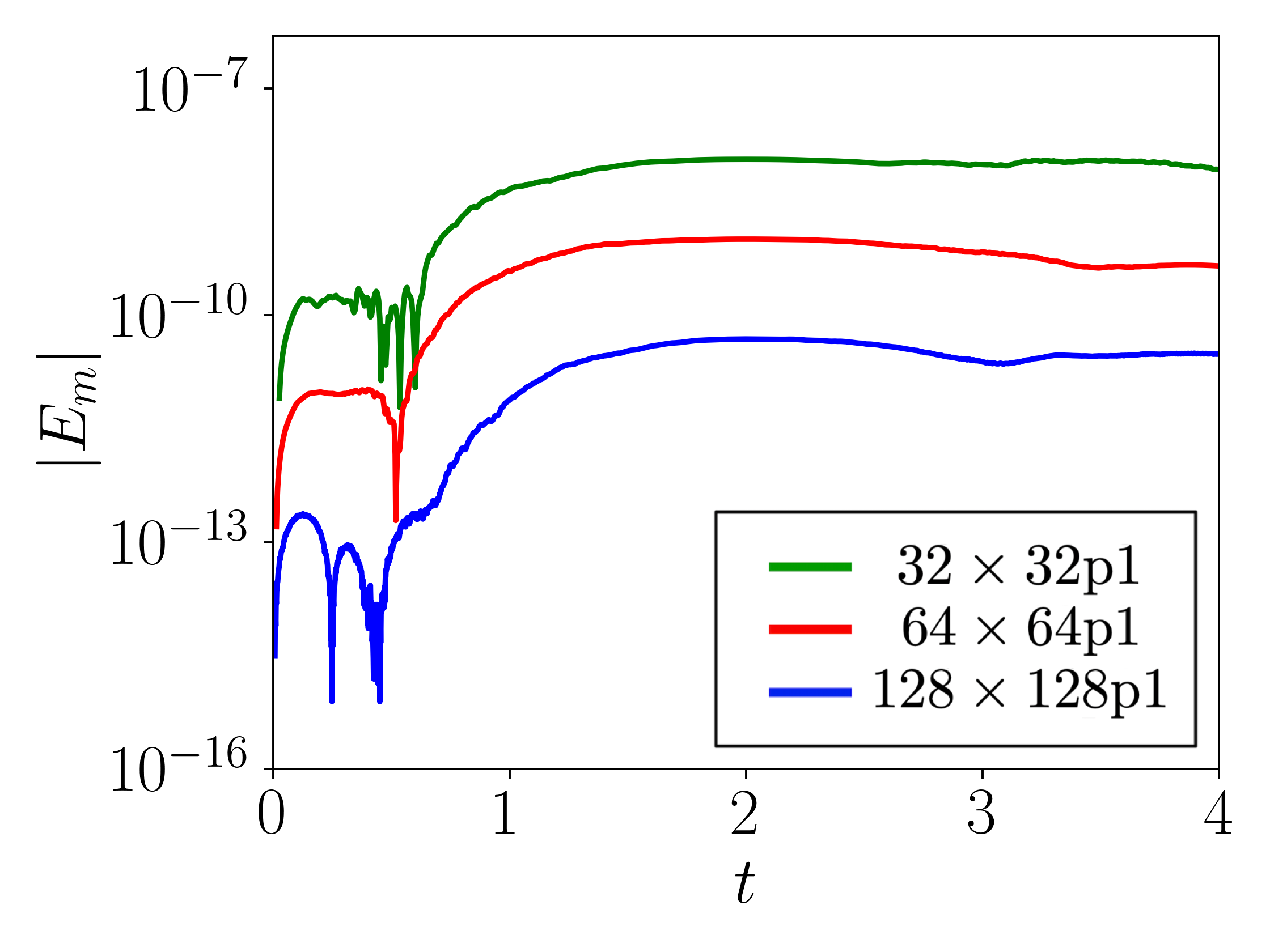}}
\subfigure{\includegraphics[width=0.49\textwidth]{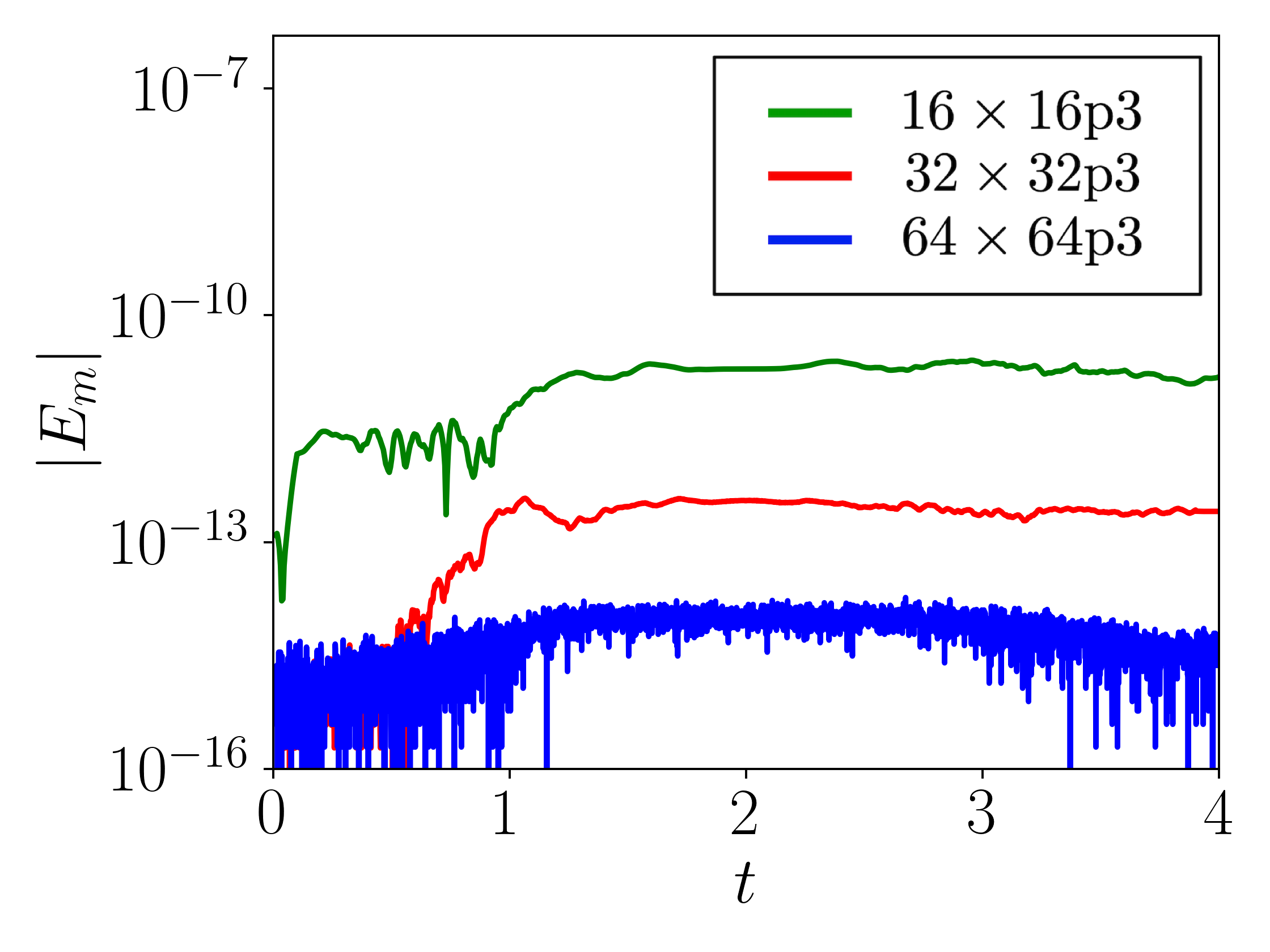}}
\caption{Mass conservation errors for different resolutions on mesh (a) using a p$1$ (left) and p$3$ approximation (right) with $64^{2}$, $128^{2}$ and $256^{2}$ DoF.}
\label{fig:mass_err}
\end{figure}

\begin{table}
 \centering
 \caption{$L^{1}$-errors for the Rider-Kothe test case compared with respect to simulations from the literature involving the same number of degrees of freedom.}
 \begin{tabular}{llll}
 \hline
 Method & $64^{2}$& $128^{2}$& $256^{2}$ \\
  \hline
 Mirjalili et al. \cite{mirjalili2020conservative}, mesh (a)   & $1.96 \times 10^{-2}$& $5.43 \times 10^{-3}$& $1.25 \times 10^{-3}$ \\
Al-Salami et al. \cite{al2021high}, mesh (a), p$3$  & $-$& $5.50 \times 10^{-3}$& $1.05 \times 10^{-3}$ \\
Present mesh (a), p$3$ & $1.75 \times 10^{-2}$& $4.83 \times 10^{-3}$& $1.23 \times 10^{-3}$ \\
Present mesh (a), p$1$ & $2.22 \times 10^{-2}$& $8.71 \times 10^{-3}$& $3.31 \times 10^{-3}$ \\
 \hline
Al-Salami et al. \cite{al2021high}, mesh (b), p$3$  & $-$& $5.51 \times 10^{-3}$& $1.25 \times 10^{-3}$ \\
Present mesh (b), p$3$ & $1.58 \times 10^{-2}$& $4.03 \times 10^{-3}$& $1.18 \times 10^{-3}$ \\
Present mesh (b),  p$1$ & $1.99 \times 10^{-2}$& $8.23 \times 10^{-3}$& $2.65 \times 10^{-3}$ \\
 \hline
 \end{tabular}
 \label{tab:L1RK}
 \end{table}
Finally, another feature of great interest for interface capturing techniques is mass conservation. Consequently, the relative mass conservation error was evaluated over time as
\begin{equation}
E_{m} = \frac{\int (\phi(\textbf{x},t) - \phi(\textbf{x},0)) d \Omega}{\int \phi(\textbf{x},0) d \Omega }.
\end{equation}
The time evolution of the mass conservation error throughout the whole simulation is shown in figure \ref{fig:mass_err} for different mesh resolutions and different orders of approximation.

Since the transport term in the phase field equation is discretised in a non-conservative form, mass conservation errors are not identically zero. 
Considering the p$1$ simulation, the mass conservation errors are relatively small, ranging between  $10^{-8}$ and  $10^{-11}$ for different mesh resolutions, as the grid size decreases. 
Even if these errors are already considerably small for a first order polynomial approximation, it can be observed that the mass conservation errors are even smaller for the p$3$ simulation, as they range between $10^{-10}$ for the coarsest mesh to almost machine precision for the most refined simulation. 

In the work by Al-Salami et al. \cite{al2021high}, in the same case for $256^{2}$ DoF, mass conservation errors were larger by $11$ orders of magnitude.
%
\subsection{Two-phase flows}
Following the analysis of the  interface capturing technique, we will considered a series of validation tests involving the simulation of two-phase flows using the five equation model presented in the previous section. 
\subsubsection{Advection of a water droplet in air}
A popular test case to validate the present implementation consists in the advection of a water droplet in air. The one-dimensional form of the present test case has been extensively used as a benchmark validation problem in the multiphase flows community \cite{saurel1999simple,allaire2002five,johnsen2006implementation,saurel2009simple}.

The goal of this test is to assess the ability of the numerical scheme to preserve the initial circular shape of the droplet for long time integration and to quantify the robustness of the solver in dealing with high-density-ratio interfaces.

Another goal of the water bubble advection case consists in verifying the capability of the numerical discretisation to exactly preserve isolated material interfaces. A well-balanced discretisation of the five equation model should, in fact, maintain pressure and velocities exactly constant during time integration. 
For this specific case viscous, gravitational and surface tension effects are neglected. 

The bubble of radius $R=25/89$ is located at the center of a $[0,1]^{2}$ periodic square. 
The material properties for the air medium for this test case are $\gamma_{1}=1.4$, $\rho_{1} = 1 \times 10^{-3}$ and $P_{1}^{\infty} = 0 $, whereas for the water medium the fluid properties are $\gamma_{2}=4.4$, $\rho_{2} = 1$ and $P^{\infty}_{2} = 6 \times 10^{3}$. 

The initial conditions read:
\begin{equation}
\textbf{u} = (5,5), \quad P=1, \quad \phi_{1} = \frac{1}{2}\bigg(1 + \tanh \bigg( \frac{r-R}{2\epsilon}\bigg) \bigg) \quad \mathrm{and} \quad \rho = \rho_{2} + (\rho_{1}-\rho_{2}) \phi_{1}.
\end{equation}

The simulation is carried on different mesh resolutions for a total of five advection periods. $3^{\mathrm{rd}}$ and $1^{\mathrm{st}}$ polynomial order DG discretisations were considered and time integration was performed using a $4^{\mathrm{th}}$ order Runge-Kutta scheme.

\begin{figure}[h!]
\centering
\includegraphics[width=.95\textwidth]{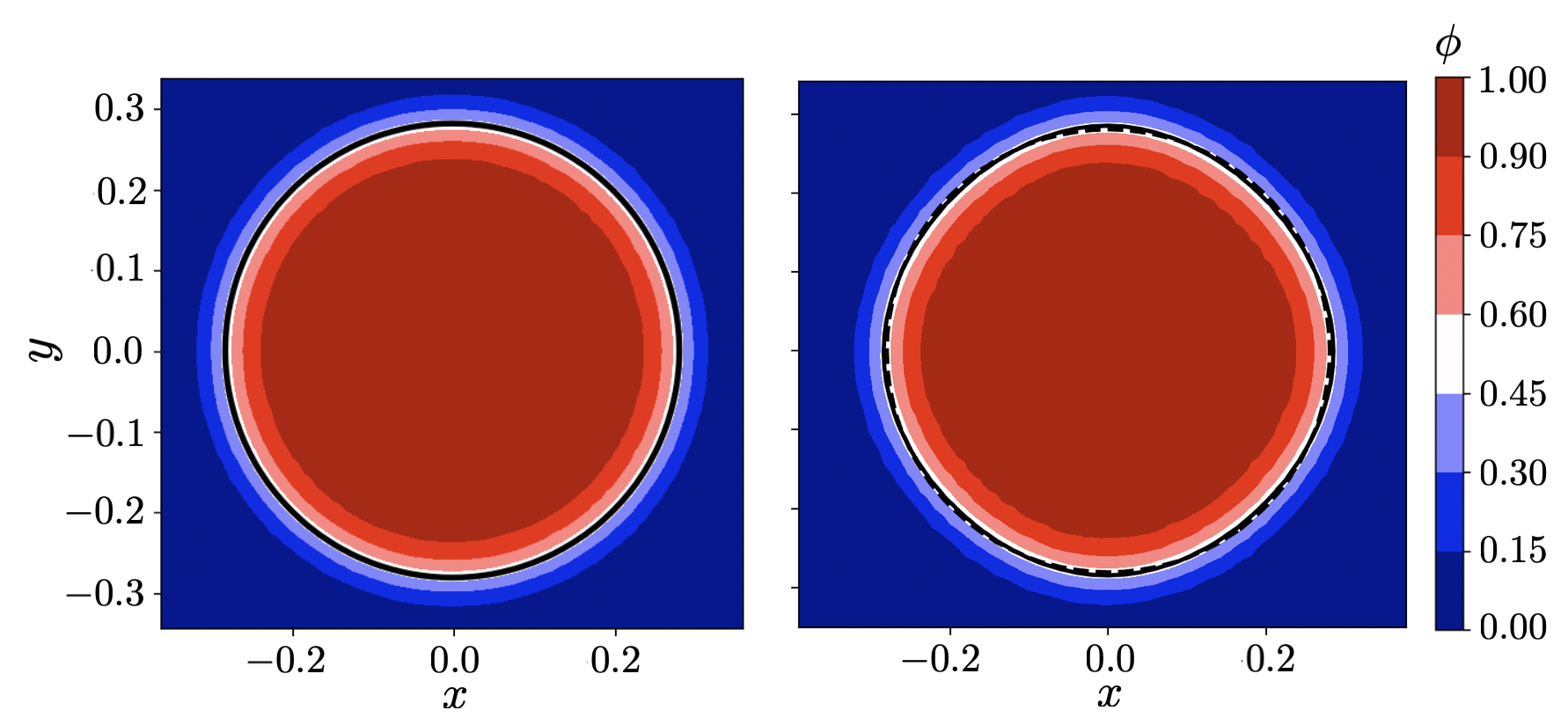}
\caption{phase field isocontours after five advection periods for the most refined simulation. Left, $24^{2}$p$3$; right, $48^{2}$p$1$. The interface is represented by the solid black line. The initial condition is shown as a dashed black line.}
\label{fig:drop_fin}
\end{figure}

In figure \ref{fig:drop_fin}, the phase field at the end of the simulations is shown for the most refined simulations (equivalent to $96^{2}$ DoF). It can be noticed that in the p$3$ simulation the droplet is qualitatively identical with respect to the prescribed initial condition, indicating that the proposed methodology is successfully able to preserve the initial shape of the water bubble even after long time integration. It is worthwhile mentioning that a spurious deformations of the interface and the grid directions is often encountered for low-order approximations of the interface normals \cite{tiwari2013diffuse}. It is known that the use of high-order methods in the advection of the phase field can mitigated such numerical artefacts \cite{tiwari2013diffuse,aslani2018localized}. The present simulation confirms this behaviour, where the p$1$ discretisation shows a much more evident deformation of the droplet after five advection periods. 

\begin{figure}[h!]
\centering
\subfigure{\includegraphics[width=0.48\textwidth]{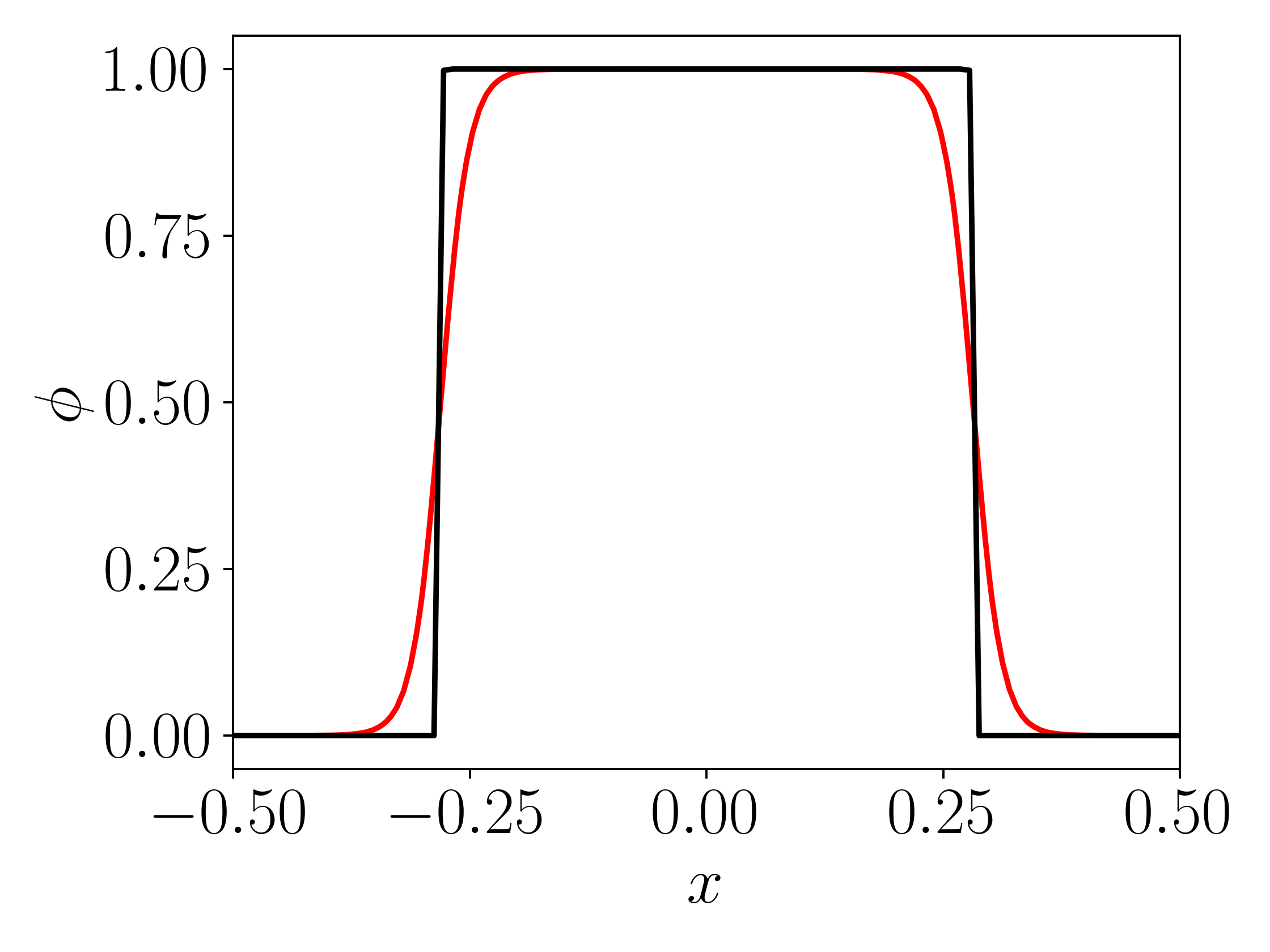}}
\hfill
\subfigure{\includegraphics[width=0.48\textwidth]{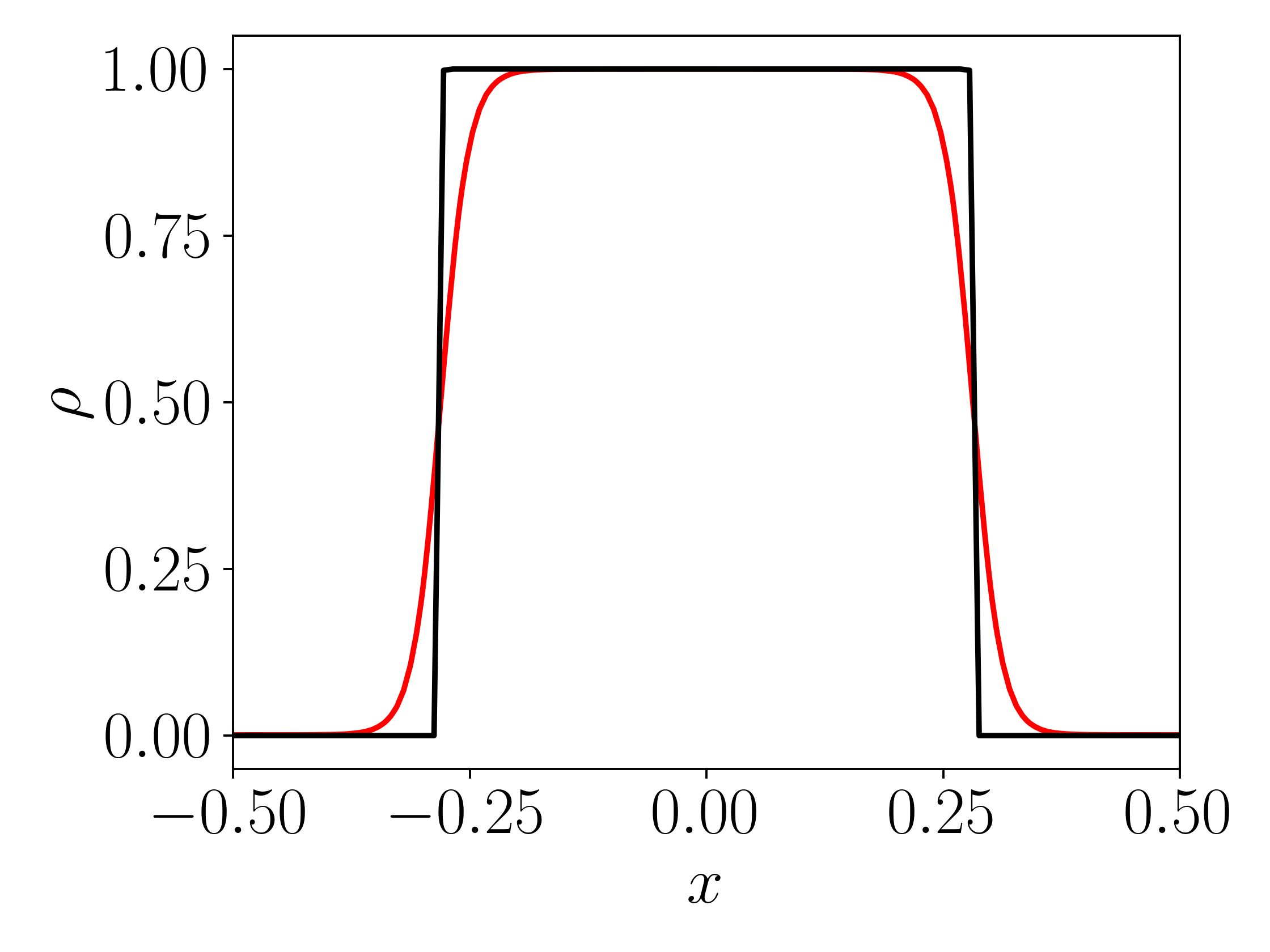}}
\vfill
\subfigure{\includegraphics[width=0.48\textwidth]{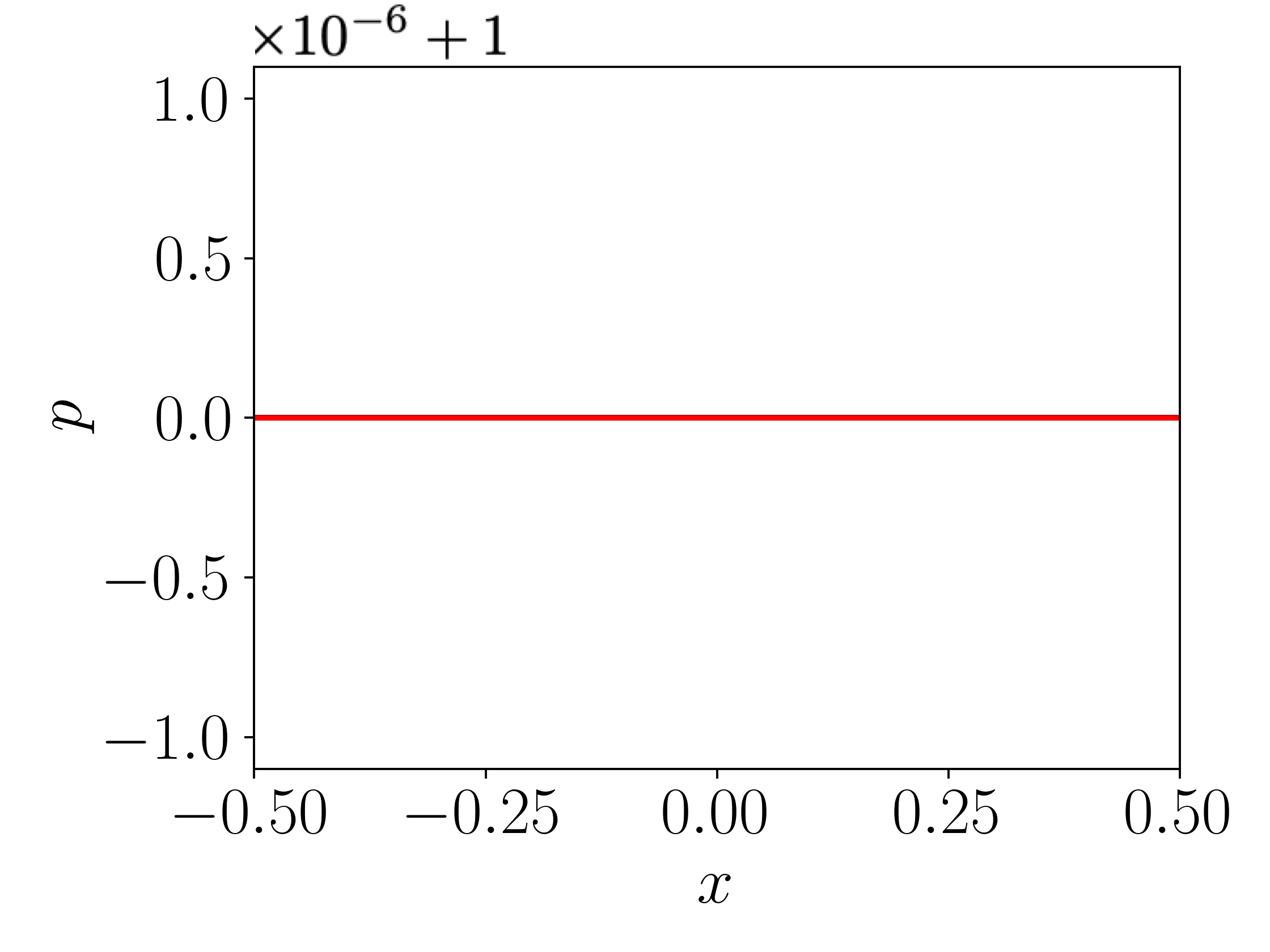}}
\hfill
\subfigure{\includegraphics[width=0.48\textwidth]{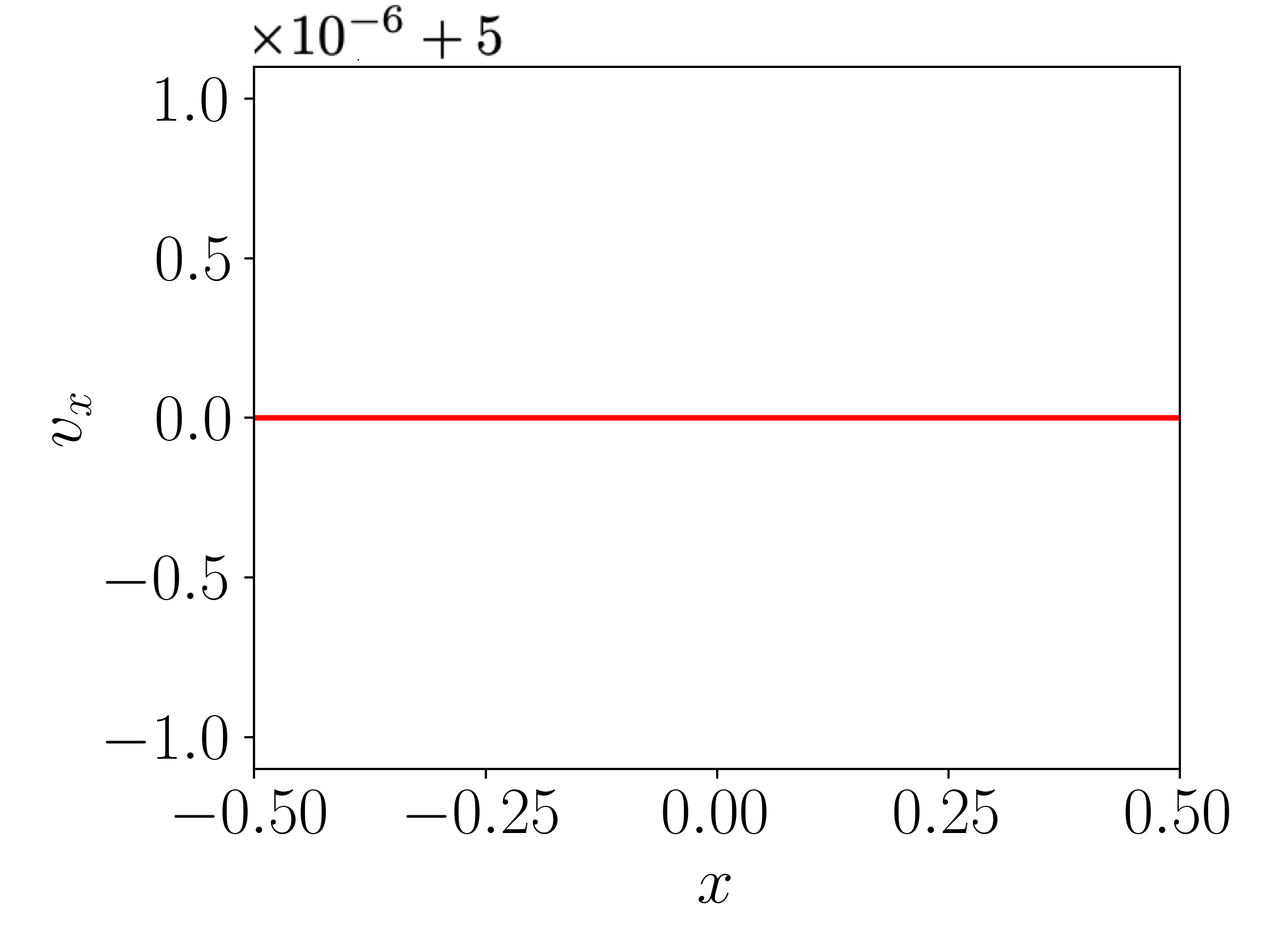}}
\caption{Top row: phase field (left) and density (right) along the line $y=0$ after five advection periods. Bottom row: slice of pressure (left) and $x$-velocity (right) along the line $y=0$ after five advection periods. Red, numerical simulation; black, exact solution. Notice that the $y$-axis is centered around the expected exact value and scaled by $10^{-6}$.}
\label{fig:slice}
\end{figure}

These results further confirm the overall benefit of using high-order discretisations already observed in the previous Rider-Kothe test case.
In order to have a better understanding of the overall behaviour of all the relevant quantities, in figure \ref{fig:slice} the phase field, density, pressure and $x$-velocity are respectively plotted along the line $y=0$ for the p$3$  most refined simulation ($24^{2}$p$3$). It can be observed that both the phase field and the total density vary smoothly across the interface, preserving the prescribed hyperbolic tangent profile without any spurious oscillation in proximity of the interface. 

Finally, we remark that pressure and $x$-velocity remain uniform during the simulation as shown in figure \ref{fig:slice}, indicating that the proposed discretisation is able to fulfil the interface equilibrium condition.
%
%
\subsubsection{Rayleigh-Taylor instability}
The Rayleigh–Taylor instability occurs when an interface between two fluids with different densities experiences a pressure gradient opposing the density gradient. The domain consists in a $[d \times 4d ]$ with $d=1$. The interface is initially defined as the curve $y(x) = 2d + 0.1 d \cos{(2\pi x/d)}$. The Rayleigh-Taylor instability is characterised by the Reynolds number $\mathrm{Re} = (\rho_{1} d^{3/2} ||\textbf{g}||^{1/2})/\mu$ and the Atwood number $\mathrm{At} = (\rho_{1} - \rho_{2})/(\rho_{1} + \rho_{2})$ which are respectively set to $3000$ and $0.5$.
The top boundary is treated as a Riemann-invariant boundary condition with zero velocity and constant pressure, the bottom boundary is a no-slip wall whereas slip wall boundary conditions are prescribed on the other lateral sides of the domain.  
Two different quadrilateral grids were considered for this study involving $5000$ and $20000$ degrees of freedom ($10\times20$p$4$ and $20\times80$p$4$, respectively). Finally, a $4^{\mathrm{th}}$ order Runge-Kutta scheme was used for time integration.

The present test is meaningful in considering a more complex physical set-up where viscous and gravitational force drive the dynamics of the system.

\begin{figure}[h!]
\centering
\subfigure{\includegraphics[width=0.95\textwidth]{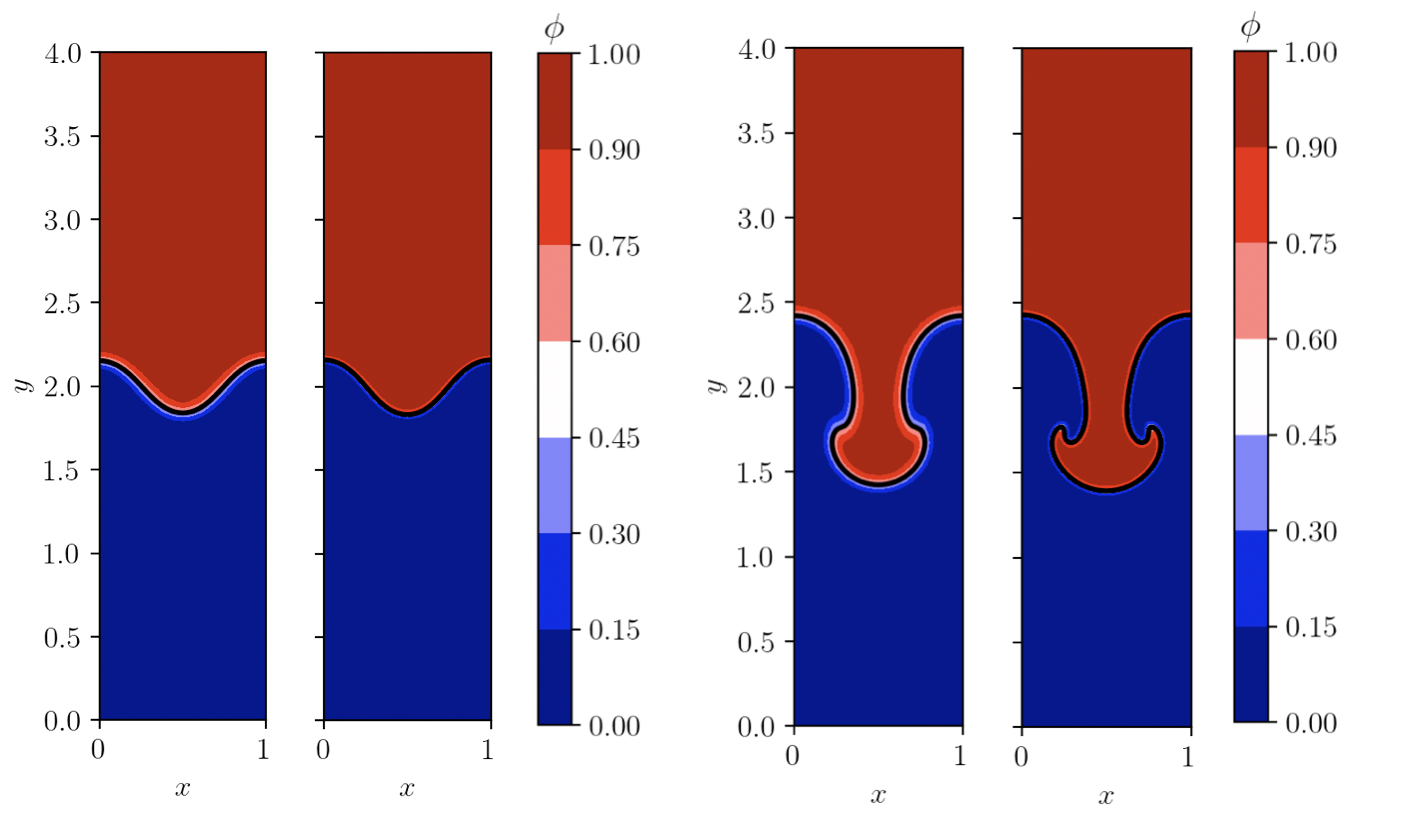}}
\vfill
\subfigure{\includegraphics[width=0.95\textwidth]{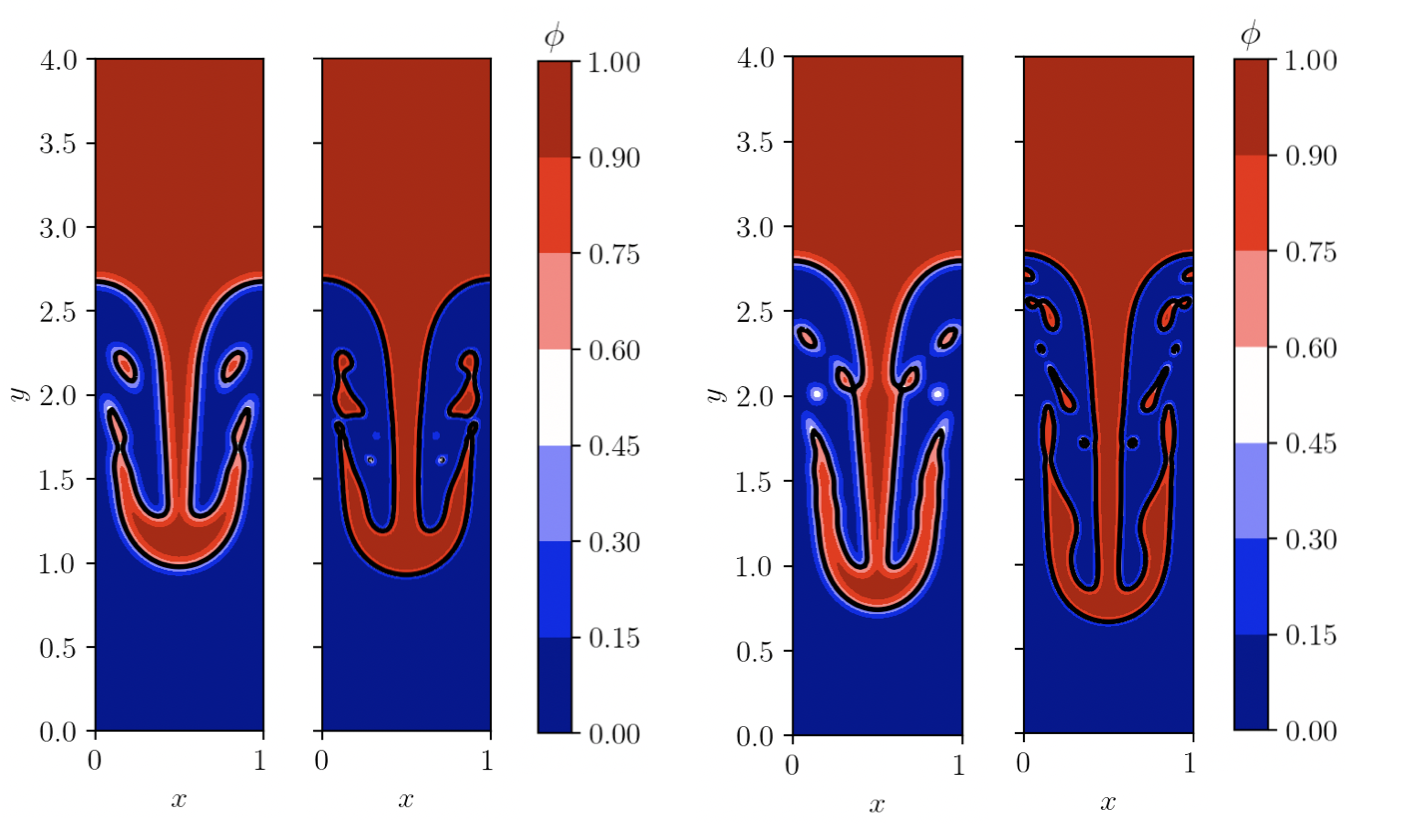}}
\caption{Snapshots of the phase field isocontours for different meshes and times. Top row: isocontour of the phase field at $t^{*}=0.5$ (left) and $t^{*}=1.5$ (right). Bottom row: isocontour of the phase field at $t^{*}=2.5$ (left) and $t^{*}=3.0$ (right). In each subfigure: left, $10\times20$p$4$; right $20\times80$p$4$. The interface is highlighted in black.}
\label{fig:RT_snap}
\end{figure}

A series of snapshots of the phase field are shown in figure \ref{fig:RT_snap} for the two grids. 
Clearly, the flow field arising from the Rayleigh-Taylor instability is much richer than the previous test case: viscous and gravitational forces lead to complex vortical structures causing non-trivial dynamics of the interface, including primary breakup. These structures are increasingly better resolved under mesh refinement as it can be observed from figure \ref{fig:RT_snap}. 

\begin{figure}[h!]
\centering
\includegraphics[width=.75\textwidth]{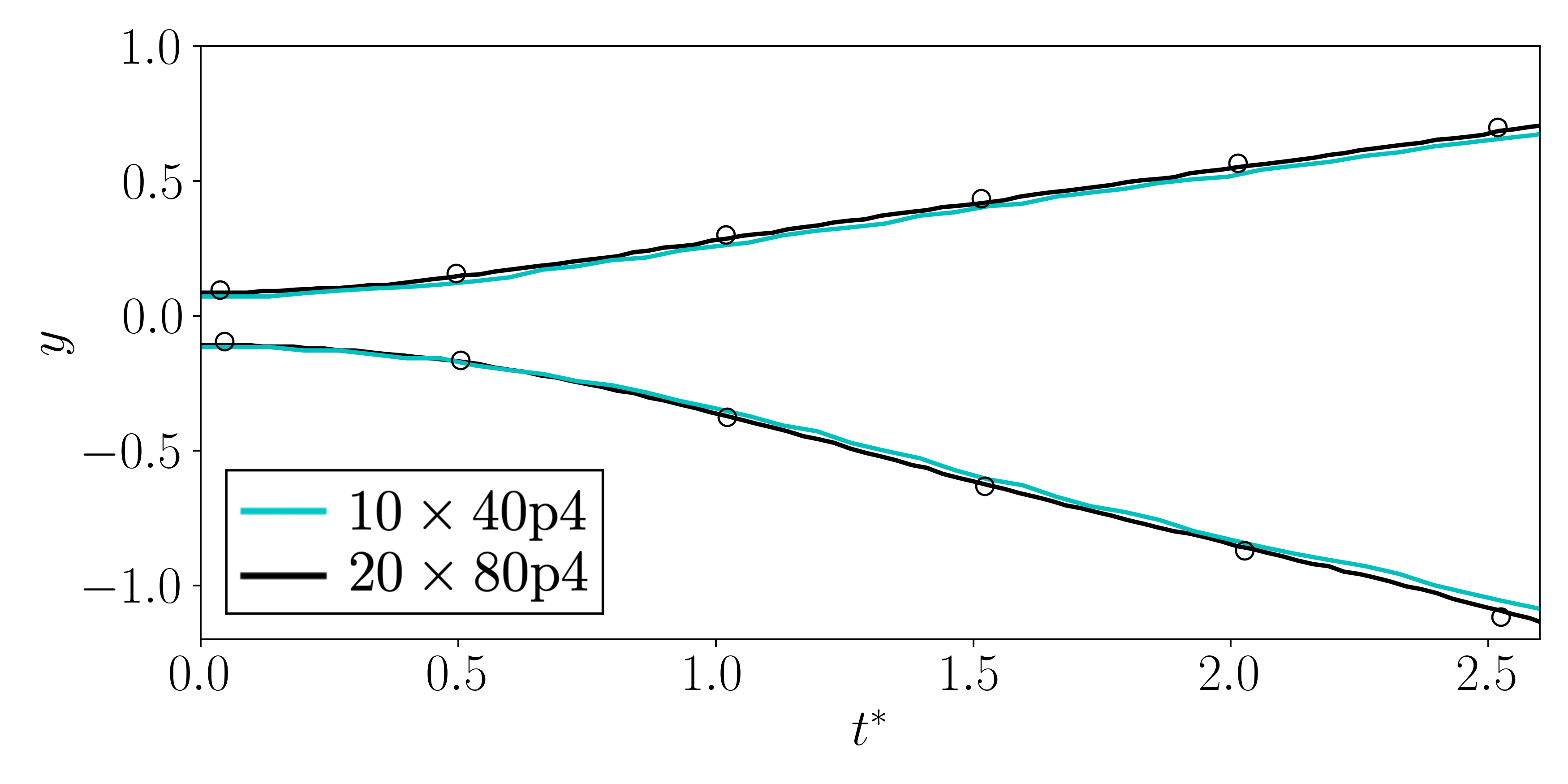}
\caption{The evolution of the top and bottom of the interface for the Rayleigh-Taylor instability versus non-dimensional time for $10\times40$p$4$ and $20\times80$p$4$. Symbols indicated the reference solution by Chiu \& Lin \cite{chiu2011conservative}.}
\label{fig:plumes}
\end{figure}

From a more quantitive point of view, the predicted top and bottom locations of the interface versus the non-dimensional time ($t^{*} = t /\sqrt{d/(||\textbf{g}|| \mathrm{At})}$) are shown in figure \ref{fig:plumes}. Excellent agreement with previous studies can be observed. 

\begin{figure}[h!]
\centering
\subfigure{\includegraphics[width=0.95\textwidth]{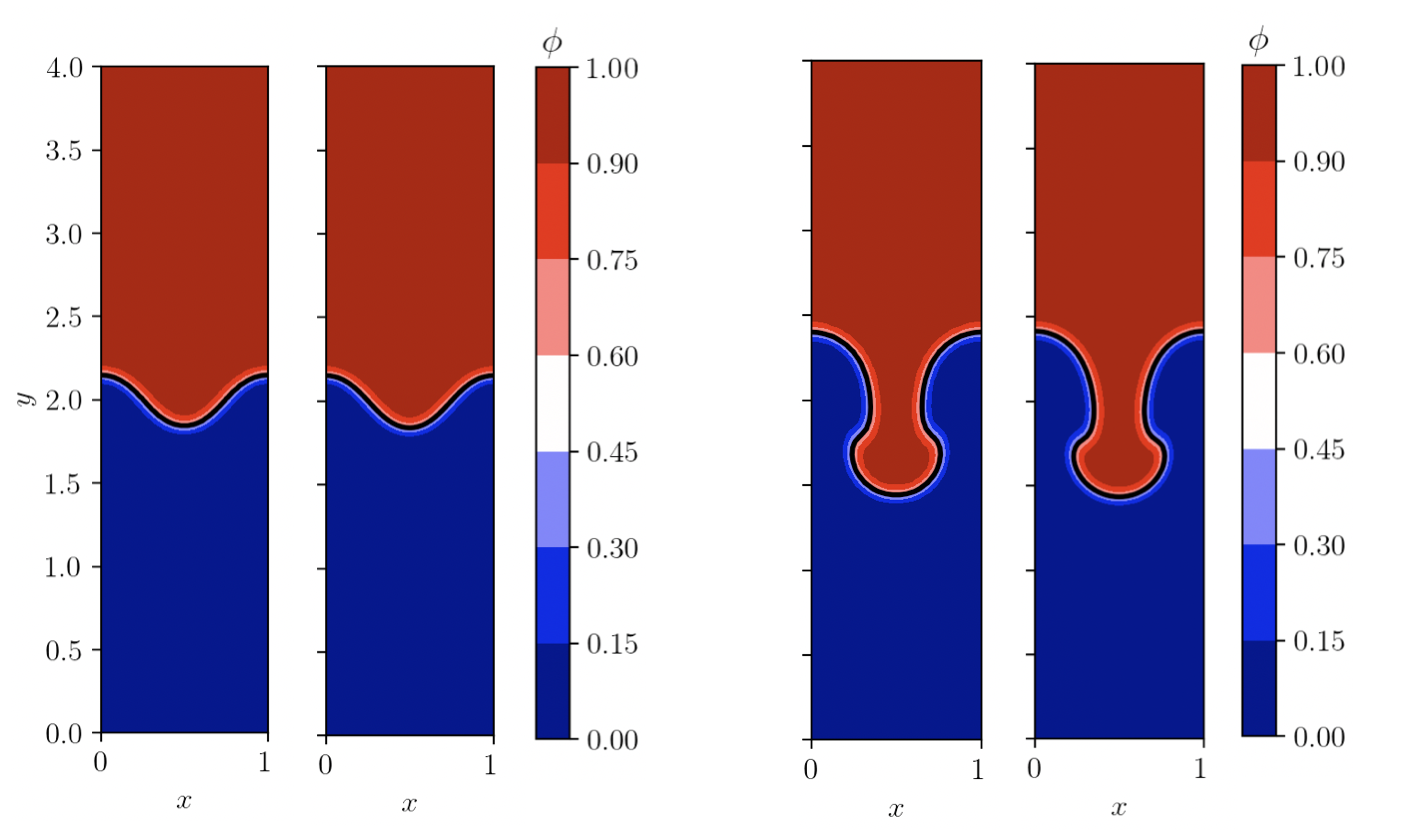}}
\vfill
\subfigure{\includegraphics[width=0.95\textwidth]{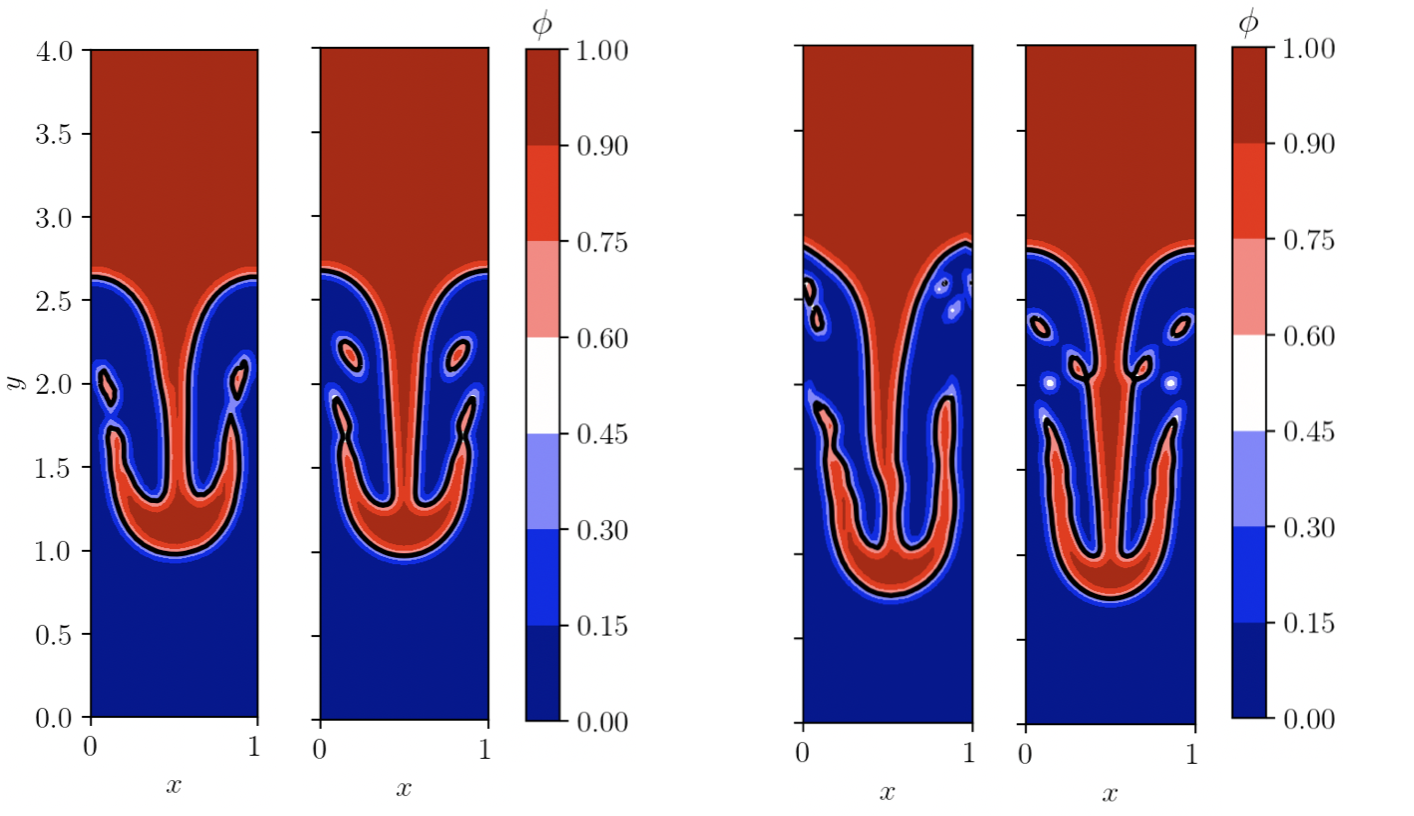}}
\caption{Snapshots of the phase field isocontours for different orders and times. Top row: isocontour of the phase field at $t^{*}=0.5$ (left) and $t^{*}=1.5$ (right). Bottom row: isocontour of the phase field at $t^{*}=2.5$ (left) and $t^{*}=3.0$ (right). In each subfigure: left, $25\times100$p$1$; right, $10\times40$p$4$. The interface is highlighted in black.}
\label{fig:RT_snap2}
\end{figure}

Similarly to the Rider-Kothe test case, the influence of the order of approximation was investigated. In particular, a $1^{\mathrm{st}}$ polynomial order simulation has been performed and compared with the $4^{\mathrm{th}}$  polynomial order computation. A series of snapshots of the phase field are shown in figure \ref{fig:RT_snap2} for the two different orders of approximation.

\begin{figure}[h!]
\centering
\includegraphics[width=.99\textwidth]{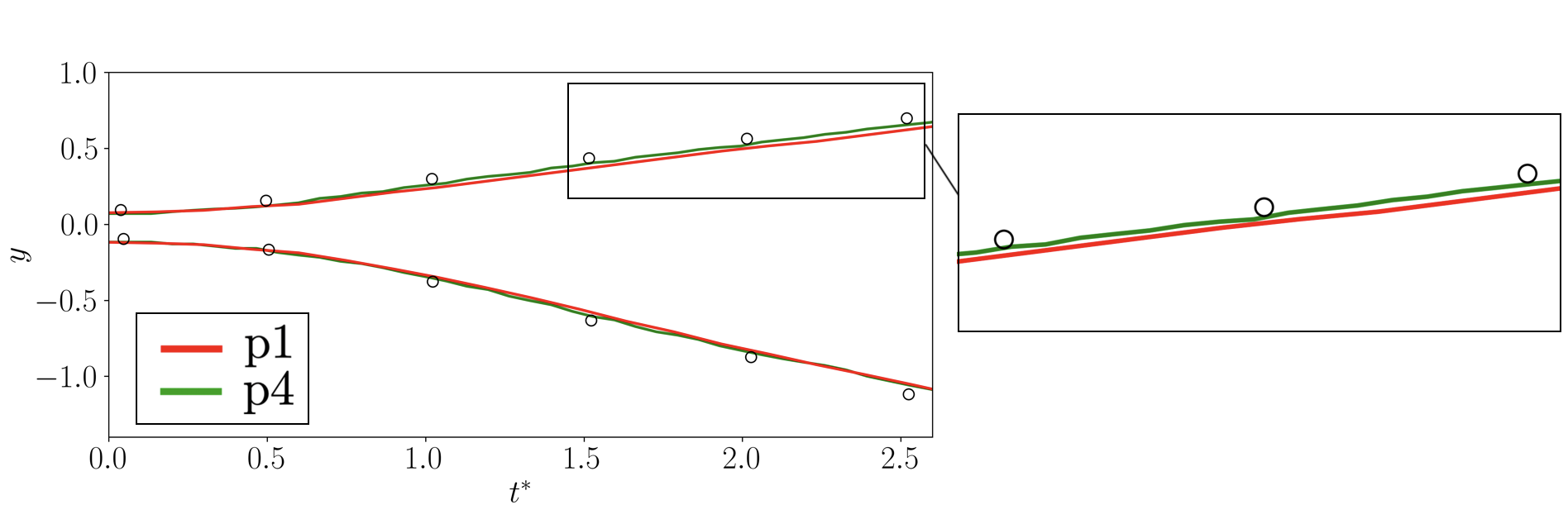}
\caption{The evolution of the top and bottom of the interface for the Rayleigh-Taylor instability versus non-dimensional time using a p$1$ ($50\times200$p$1$) and p$4$ ($20\times80$p$4$) approximation. Symbols indicate the reference solution by Chiu \& Lin \cite{chiu2011conservative}.}
\label{fig:plumes_p}
\end{figure}

\begin{figure}[h!]
\centering
\subfigure{\includegraphics[trim={0 0 0 0}, clip, width=0.6\textwidth]{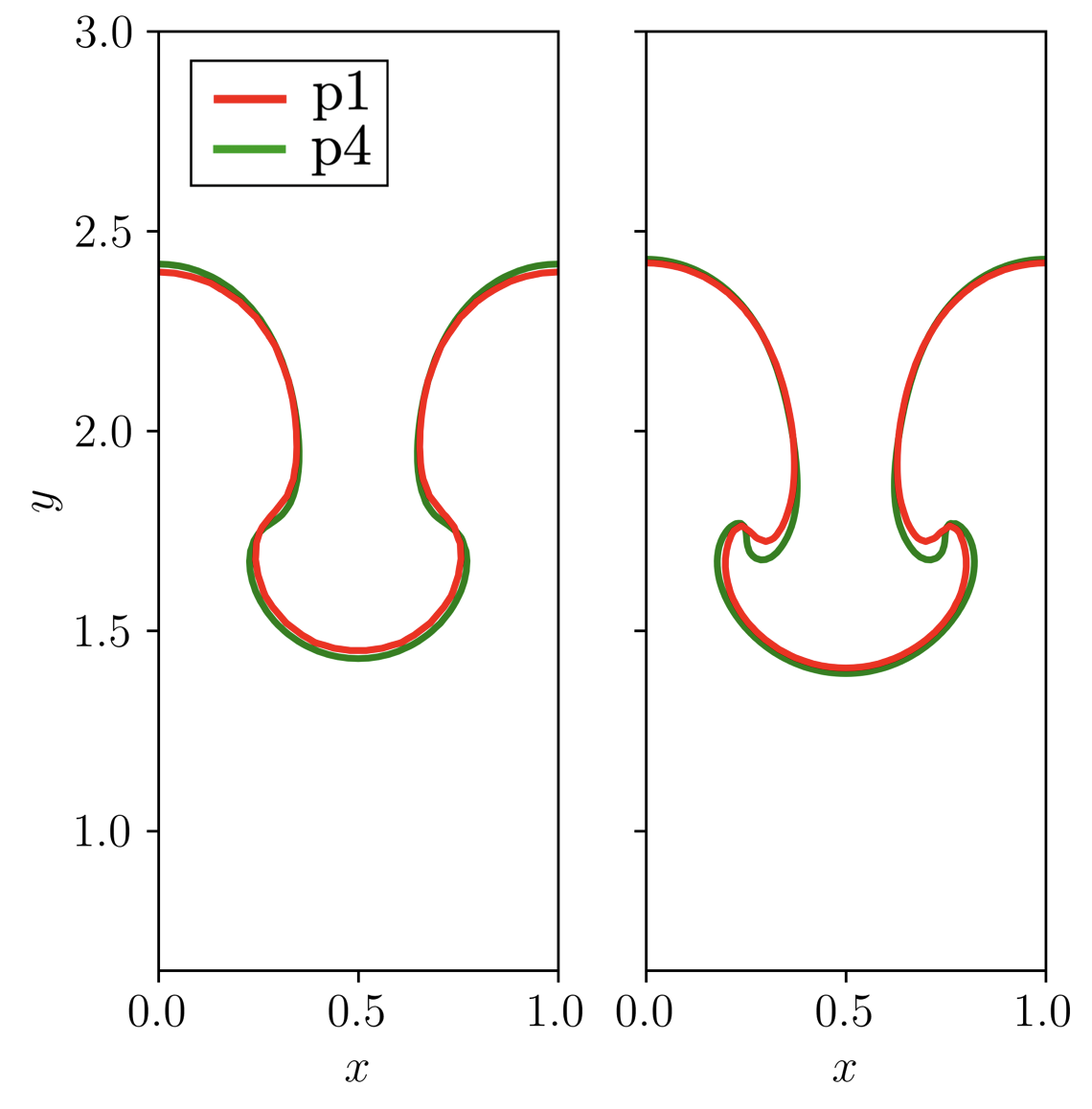}}
\caption{Iso-contour $\phi =0.5$ with $1^{\mathrm{st}}$ and $4^{\mathrm{th}}$ polynomial order approximations at $t^{*}=1.5$. Left, $50\times200$ DoF; right $100\times400$ DoF.}
\label{fig:p1_vs_p4}
\end{figure}

\begin{figure}[h!]
\centering
\subfigure{\includegraphics[trim={0 0 0 0}, clip, width=0.6\textwidth]{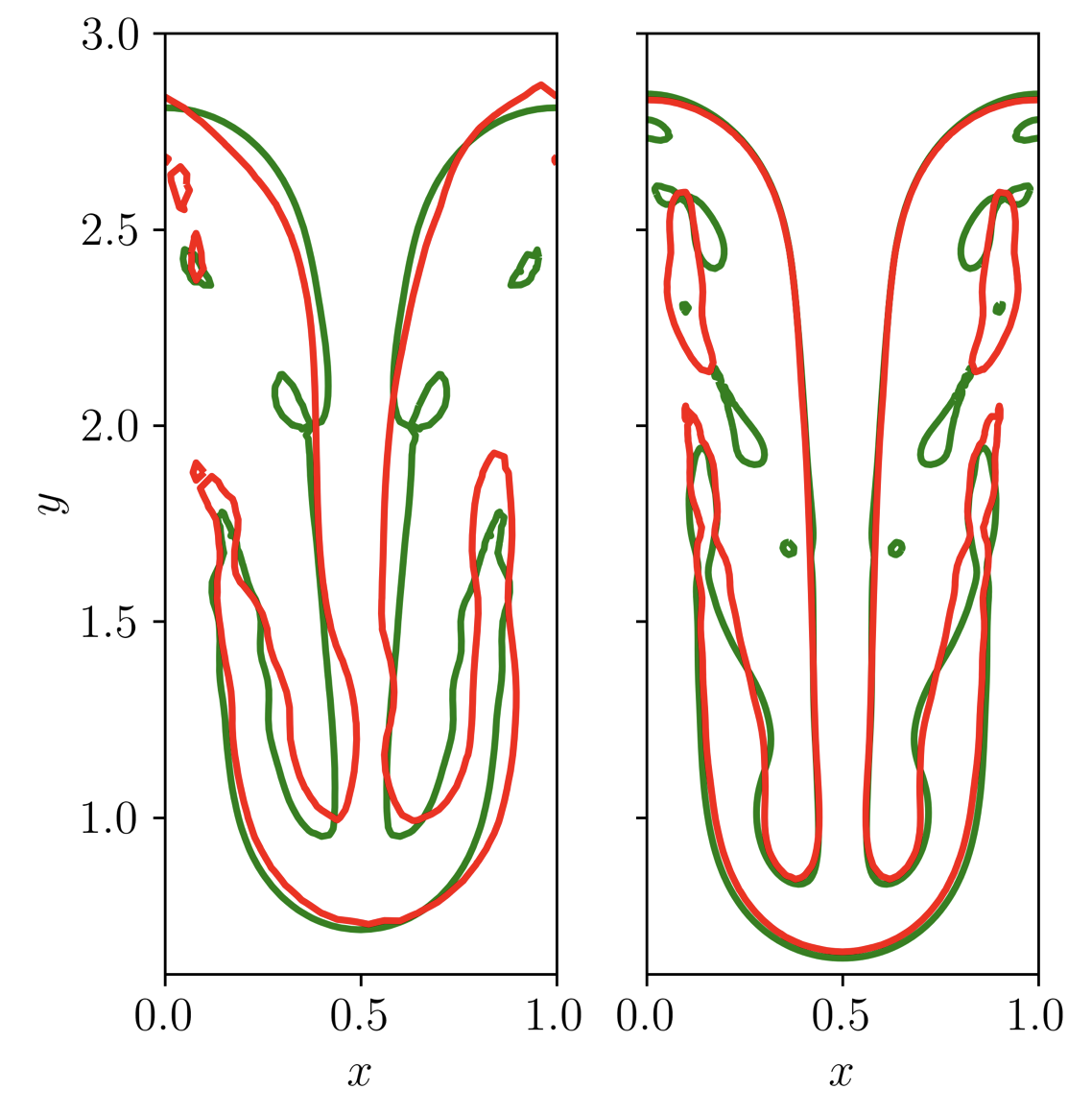}}
\caption{Iso-contour $\phi =0.5$ with $1^{\mathrm{st}}$ and $4^{\mathrm{th}}$ polynomial order approximations at $t^{*}=3.0$. Left, $50\times200$ DoF; right $100\times400$ DoF.}
\label{fig:p1_vs_p4_2}
\end{figure}

It can be noticed that the p$1$ simulation provides less accurate results, leading to an asymmetric solution. 
From a more quantitative point of view, the location of the upper and lower plumes were compared for the two different orders of approximations and are shown in figure \ref{fig:plumes_p}. It can be noticed that the p$4$ approximation is always closer to the reference solution throughout the whole simulation, in particular for the prediction of the top location of the interface.
To better appreciate the differences between high and low orders of approximation, the locations of the interface at $t^{*}=1.5$ and $t^{*}=3.0$ are shown in figures  \ref{fig:p1_vs_p4} and \ref{fig:p1_vs_p4_2} at different resolutions and different polynomial orders. 
It can be noticed that the p$1$ approximation is characterised by smoother, over-dissipated profiles along the tangential direction of the interface. Remember that the smoothness along the normal direction is governed by the interface capturing technique and it should not be particularly influenced by the spatial order of approximation, but the accuracy along the tangential direction of the interface, instead, should be more sensitive to the approximation order.
At later times, when primary beak-up occurs, more complex small scale structures are captured by the high-order simulation, in particular for the simulations on the finest grid.
It can also be noticed that the p$4$ simulation is characterised by a more stretched interface, (\ie bottom and top locations of the interface are respectively lower and higher with respect to the corresponding p$1$ simulations). This behaviour is in agreement with what is observed in figure \ref{fig:plumes_p}. 
%
\subsubsection{Rising bubble} \label{sec:rising_bubble}
In this test case the rise of a $2$D bubble of light fluid within a heavier fluid due to buoyancy is simulated. Initially, a circular bubble of radius $R=0.25$ is placed at $(0.0,0.5)$ in a $[-0.5,0.5] \times [0, 2]$ domain. The density and viscosity of the fluids are chosen such that $\rho_{1}/\rho_{2}=\mu_{1}/\mu_{2}=10$. The Reynolds number is set to $\mathrm{Re} = (\rho_{1} d^{3/2} ||\textbf{g}||^{1/2})/\mu_{1}=35$ whereas the Eötvös number is $E_{0} = 4 \rho_{1} ||\textbf{g}|| r^{2}/\sigma=10$. Similarly to the previous test case, at the upper boundary a Riemann-invariant boundary condition with zero velocity and constant pressure is imposed and the bottom boundary is a no-slip wall boundary. Left and right boundaries are treated as slip walls.

Notice that the reference solution coincides with the results presented in \cite{hysing2009quantitative}.

\begin{figure}[h!]
\centering
\includegraphics[width=.95\textwidth]{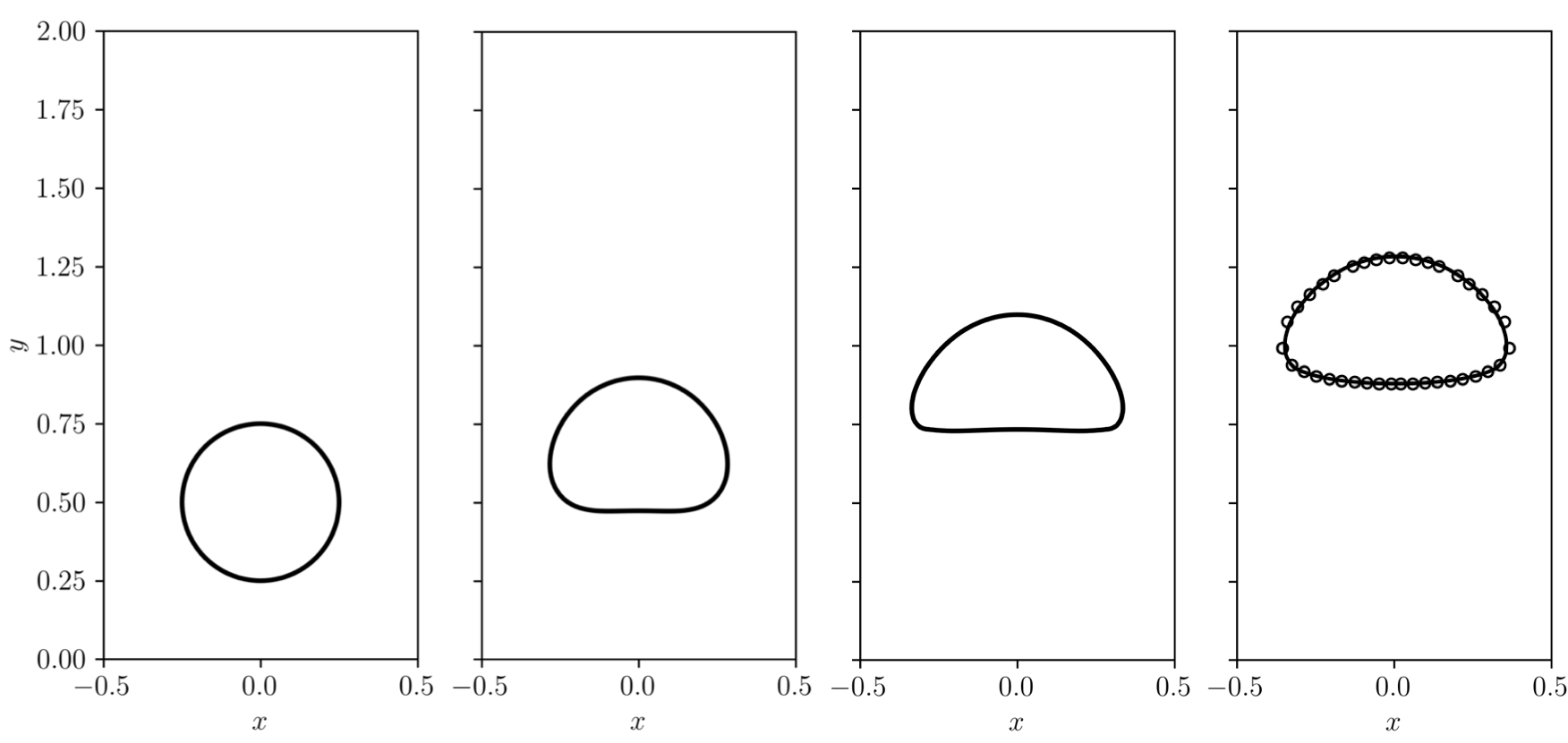}
\caption{Location of the interface at different times for the most refined simulation ($20\times40$p$4$). In order, $t=0.0,1.0,2.0,3.0$.  At final time, symbols indicate the reference solution by Hysing et al. \cite{hysing2009quantitative}.}
\label{fig:RB_snap}
\end{figure}

\begin{figure}[h!]
\centering
\subfigure{\includegraphics[width=0.48\textwidth]{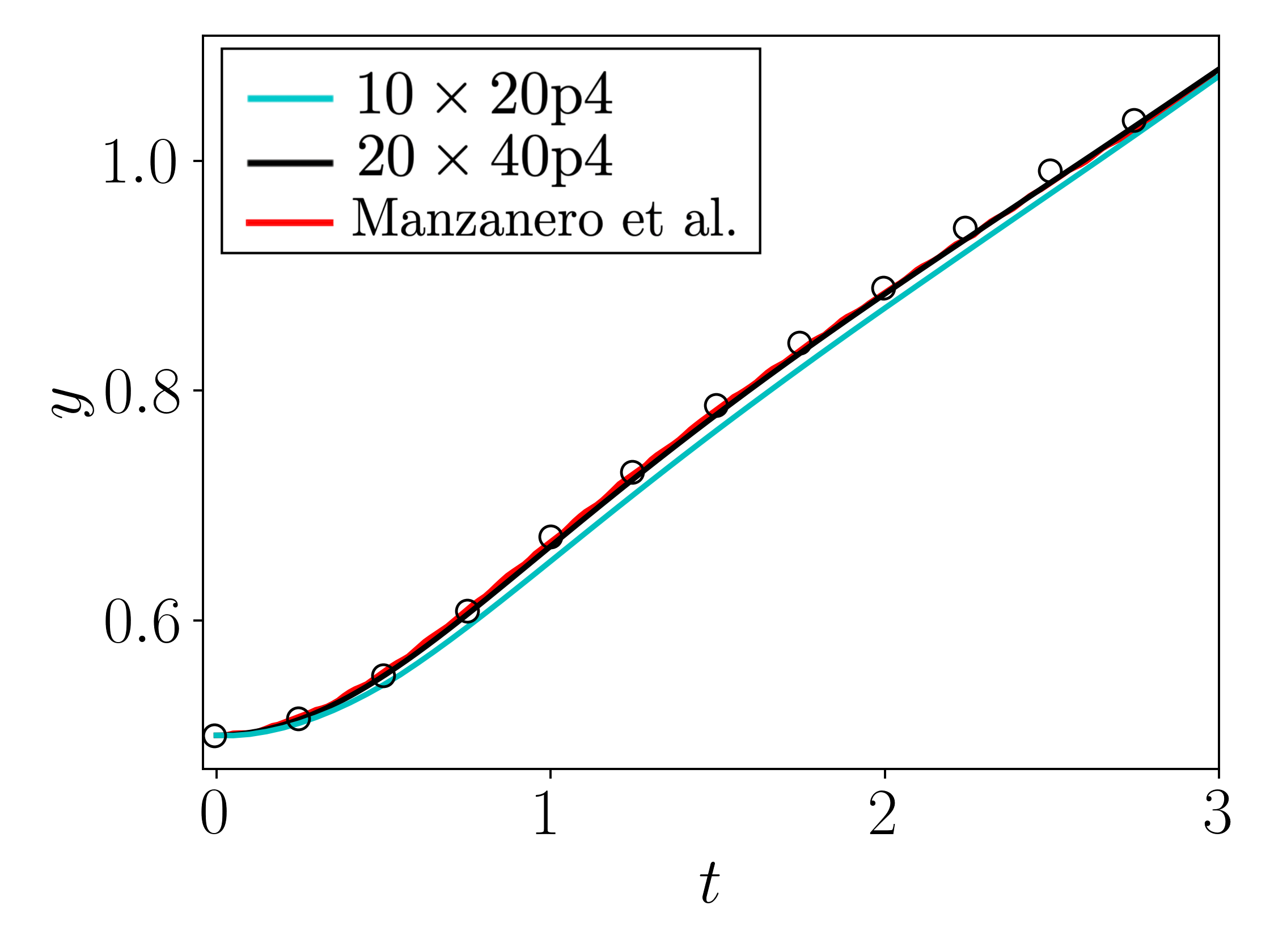}}
\hfill
\subfigure{\includegraphics[width=0.48\textwidth]{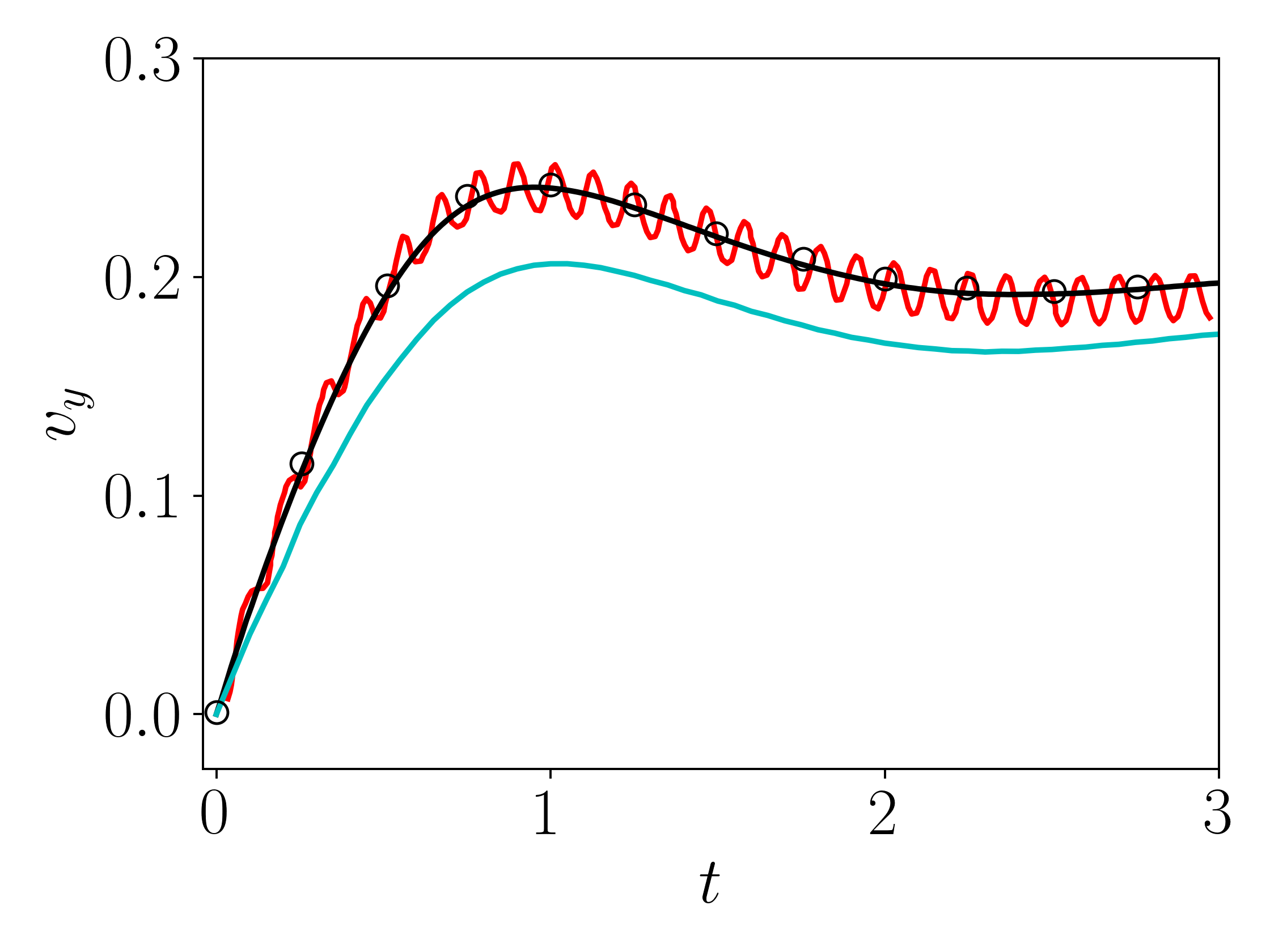}}
\caption{Location of the bubble's center of gravity (left) and bubble's vertical velocity (right) over time for the present simulation $10\times20$p$4$, $20\times40$p$4$ and results by Manzanero et al. \cite{manzanero2020entropy}.}
\label{fig:yloc}
\end{figure}

Figure \ref{fig:RB_snap} shows the evolution of the bubble in time as it rises due to buoyancy. 
Finally, the location of the center of gravity of the bubble and its vertical velocity are shown in figure \ref{fig:yloc}. In the work by Manzanero et al. \cite{manzanero2020entropy}, a similar numerical framework based on the DG method was presented. In their case, a weakly compressible approach was employed and coupled with the Cahn-Hilliard equations. It can be noticed that the pressure reflections coming from the wall boundaries cause visible oscillations in the bubble's rising velocity. Our simulation produces a similar dynamics of the bubble without spurious oscillations. In our case, a Riemann invariant boundary condition was prescribed on the upper boundary, avoiding undesirable pressure reflections.

A more challenging test case with larger density ratio and smaller surface tension coefficient was also considered. In particular, the density and viscosity of the two phases were chosen such that $\rho_{1}/\rho_{2} = 1000$ and $\mu_{1}/\mu_{2}=100$. Furthermore, the Eötvös number was increased to $E_{0} = 125$ whereas the Reynolds number remained the same. Under these conditions, the surface tension effects are not strong enough to prevent strong deformations of the bubble. 

\begin{figure}[h!]
\centering
\includegraphics[width=.95\textwidth]{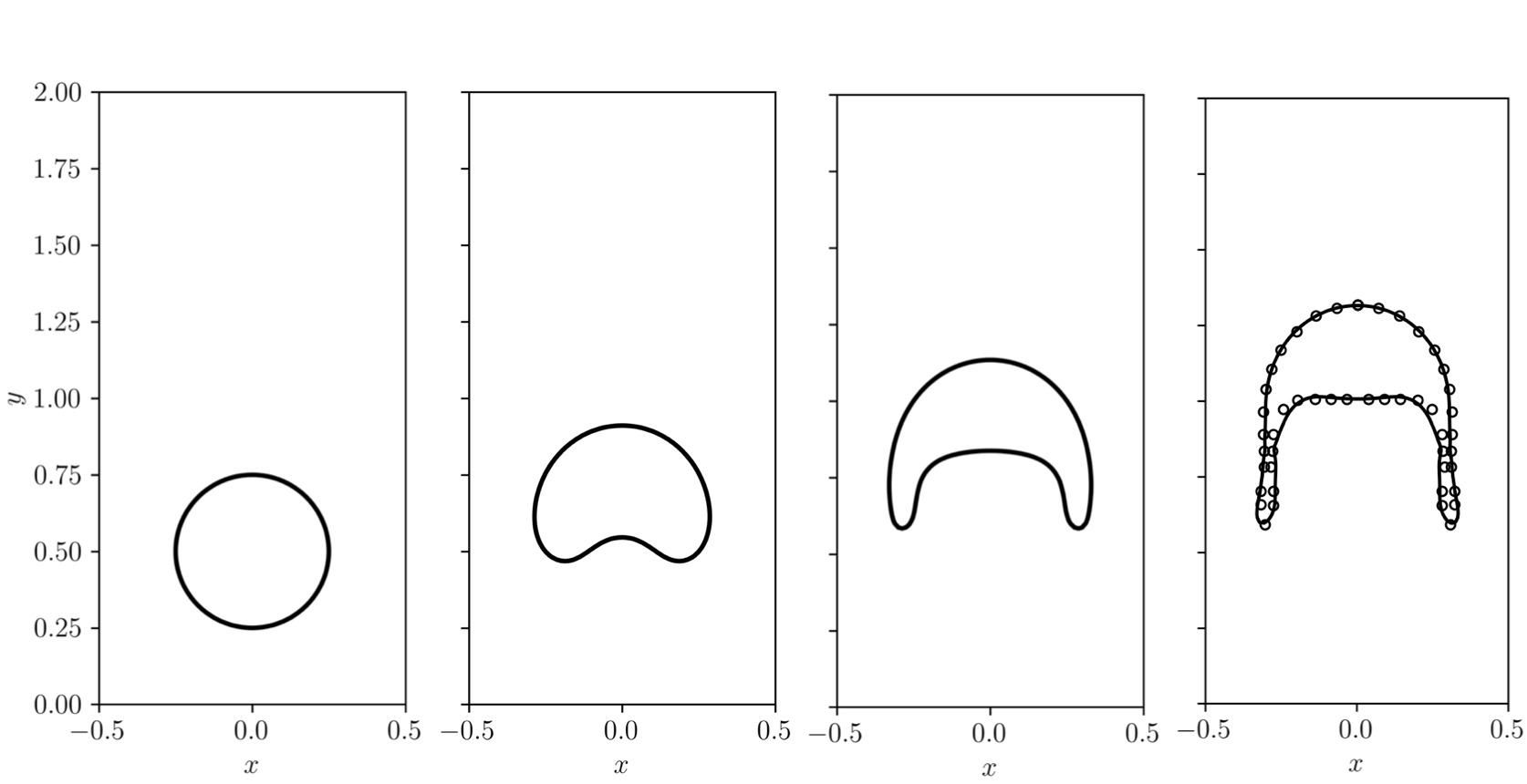}
\caption{Location of the interface at different times for the most refined simulation ($20\times40$p$4$). In order, $t=0.0,1.0,2.0,3.0$. At final time, symbols indicate the reference solution from \cite{manzanero2020entropy}.}
\label{fig:RB_snap2}
\end{figure}

In figure \ref{fig:RB_snap2}, the evolution of the bubble is shown. It can be seen that at late times, the edges of the bubble tend to elongate in considerably thin ligaments. This behaviour is caused by the insufficiently small surface tension forces with respect to gravitational forces.
Similarly with respect to the previous case, the final location of the interface was compared with a reference solution. It is well known that for this specific test case, different codes produce slightly different results in terms of the elongated ligaments. In figure \ref{fig:RB_snap2}, at the final time, the present simulation is consequently compared with the results by Manzanero et al. \cite{manzanero2020entropy}, who proposed a similar diffused-interface approach based on the Cahn-Hilliard equations. From this comparison, it can be seen that the predicted location of the interface agrees well with the reference solution.

\begin{figure}[h!]
\centering
\subfigure{\includegraphics[width=0.48\textwidth]{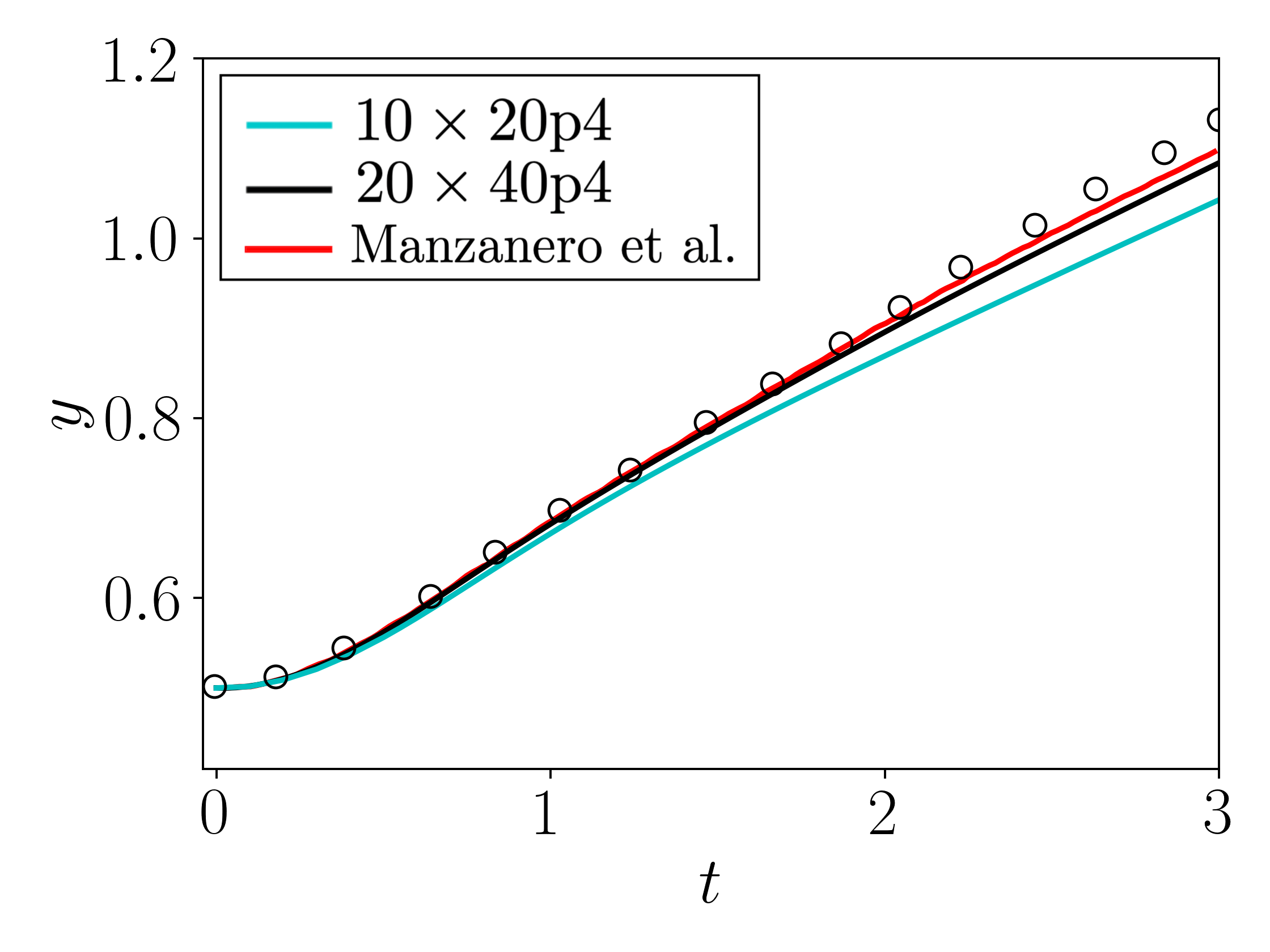}}
\hfill
\subfigure{\includegraphics[width=0.48\textwidth]{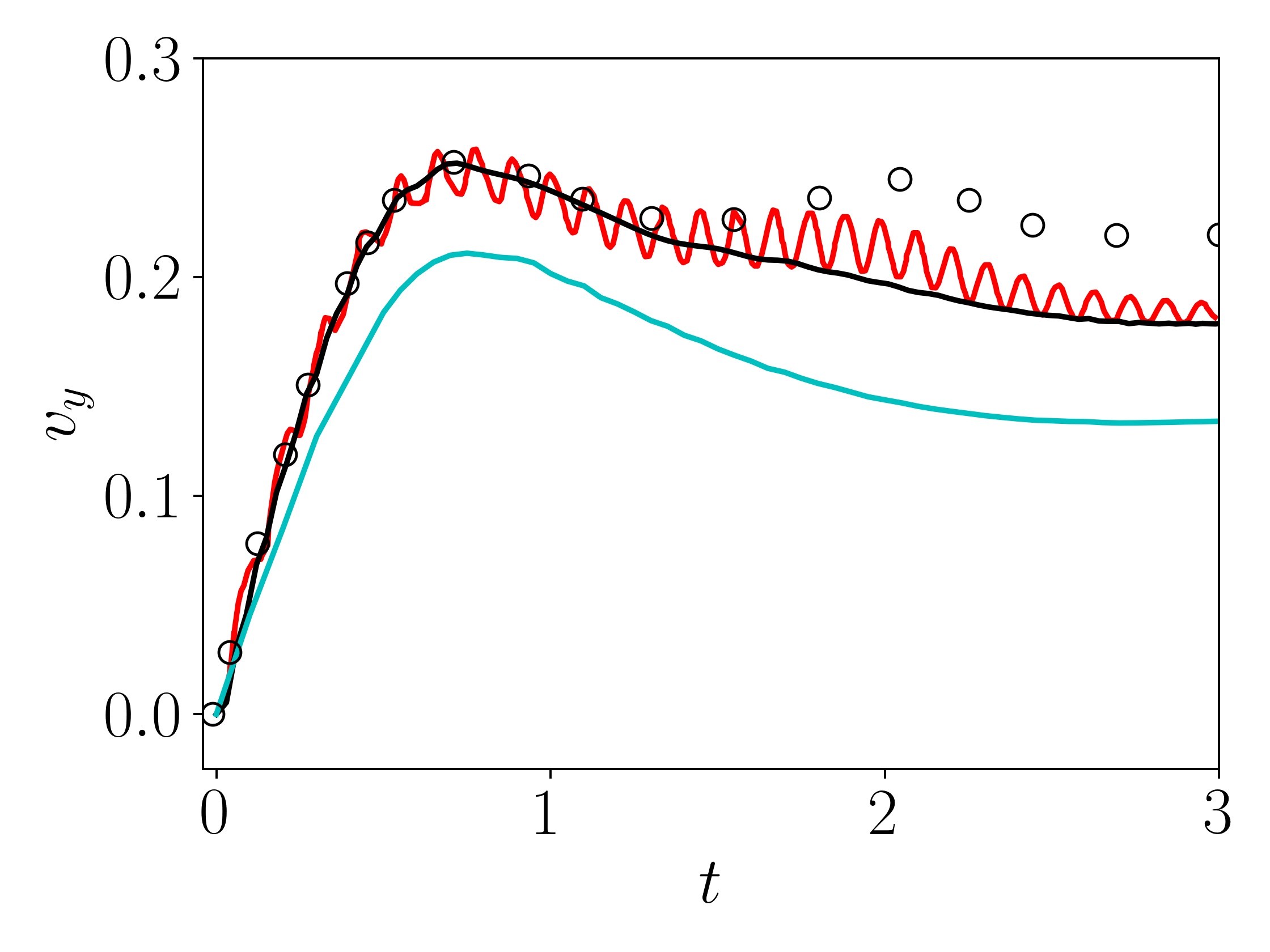}}
\caption{Location of the bubble's center of gravity (left) and bubble's vertical velocity (right) over time for the present simulation $10\times20$p$4$, $20\times40$p$4$ and results by Manzanero et al. \cite{manzanero2020entropy}.}
\label{fig:yloc2}
\end{figure}

Finally, the location of the bubble's center of gravity and the mean rising velocity of the bubble are evaluated over time and are shown in figure \ref{fig:yloc2}. The results are compared with classical reference solution from sharp interface solvers \cite{hysing2009quantitative} and the diffused-interface method from Manzanero et al. \cite{manzanero2020entropy}. It can be seen that a better agreement is obtained with this latter reference, due to the similarities between the two approaches. 
\section{Conclusions}\label{sec:conclusions}

A high-order numerical approach based on the Discontinuous Galerkin method was proposed for the simulation of two-phase flows. The interface between the two phases was modelled using the conservative Allen-Cahn equation, which was subsequently implemented within a five equation model to describe the motion of two immiscible compressible fluids. 

A series of benchmark cases were considered, including both kinematic tests involving the numerical resolution of the Allen-Cahn equation only and of fully coupled five equation model for the simulation of two-phase flows. 

In kinematic tests, studying the same approach for different orders of approximation, it was found that higher polynomial orders provided significant advantages in terms of overall accuracy, including smaller $L_{1}$ errors and mass conservation errors.
The same simulations were also performed with both quadrilateral and triangular elements without any ad-hoc modifications of the model's parameters, further underlining the robustness and flexibility of the proposed approach.

Similar observations were made in the simulation of two-phase flows using the five equation model. Lower orders of approximation were characterised by more pronounced spurious features such as deformations of the interface and artificial break-up.
Finally, more complex problems involving viscous effects, gravitational forces, and surface tension were considered. The proposed methodology successfully recovered accurate solutions in agreement with existing simulations in the literature.

Overall, the proposed methodology showed the capability of preserving fundamental features of two-phase flows such as the boundedness of the phase field, accurate computation of the interface normal vectors, small mass conservation errors and exact resolution of contact discontinuities, without, at the same time, giving up the significant benefits coming from high-order spatial discretisations. All of this is achieved with only minor modifications of the central core of the DG method, by choosing appropriate numerical fluxes and parameters of the model. At the same time, because of this, the methodology herein presented did not show the need of ad-hoc modifications for different type of elements or orders of approximation further emphasising its robustness, flexibility and generality.

\section*{Acknowledgments}
Financial support from NSF (Award 1909379) and Daikin is greatly appreciated.

\bibliography{biblio}

\end{document}